\documentclass[prd,aps,twocolumn,a4paper,showkeys,nofootinbib,showpacs]{revtex4-1}

\usepackage{subfigure}

\usepackage{graphicx,psfrag}
\usepackage{mathrsfs}
\usepackage{amsmath,amsfonts,amssymb}
\usepackage{multirow}
\usepackage{comment}

\usepackage{xspace}
\usepackage{ulem}
\usepackage{hyperref}
\usepackage{enumitem}

\newcommand{\be}{\begin{equation}}
\newcommand{\ee}{\end{equation}}
\newcommand{\bea}{\begin{eqnarray}}
\newcommand{\eea}{\end{eqnarray}}
\newcommand{\bel}{\begin{align}}
\newcommand{\eel}{\end{align}}

\newcommand{\tGRAthena}{\texttt{GR-Athena++}}
\newcommand{\GRAthena}{\tGRAthena\xspace}

\newcommand{\tTHC}{\texttt{THC}}
\newcommand{\THC}{\tTHC\xspace}

\newcommand{\tAthena}{\texttt{Athena++}}
\newcommand{\Athena}{\tAthena\xspace}

\newcommand{\orcid}[1]{\href{https://orcid.org/#1}{
\includegraphics[width=10pt]{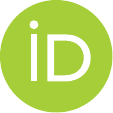}
}}

\def\Msun{{\rm M_{\odot}}}

\def\GMc2{{\rm G M_{\odot} c^{-2}}}

\def\kt2{\kappa^\text{T}_2}

\def\kt2{\kappa^\text{T}_2}

\def\2nd{2^\mathrm{nd}}
\def\4th{4^\mathrm{th}}
\def\6th{6^\mathrm{th}}
\def\8th{8^\mathrm{th}}

\def\z4c{$\mathrm{Z}4\mathrm{c}$}
\def\z4oc{$\mathrm{Z}4(\mathrm{c})$}
\def\z4{$\mathrm{Z}4$}
\def\ccz4{$\mathrm{CCZ}4$}

\def\m1{$\mathrm{M}1$}

\def\ixdi{{}_\mathrm{i}}
\newcommand{\ixdim}[1]{\ensuremath{{}_{\mathrm{i}-#1}}}
\newcommand{\ixdip}[1]{\ensuremath{{}_{\mathrm{i}+#1}}}
\def\ixdimh{{}_{\mathrm{i}-1/2}}
\def\ixdiph{{}_{\mathrm{i}+1/2}}
\def\ixdipmh{{}_{\mathrm{i}\pm1/2}}

\newcommand{\Mesh}{\texttt{Mesh}}
\newcommand{\MeshBlock}{\texttt{MeshBlock}}

\usepackage{color}
\definecolor{cyan}{rgb}{0,0.9,0.9}
\definecolor{orange}{rgb}{0.9,0.5,0}
\definecolor{purple}{rgb}{0.8,0.4,0.8}
\definecolor{grey}{rgb}{0.8242,0.8242,0.8242}
\definecolor{brickred}{rgb}{0.8, 0.25, 0.33}
\definecolor{darkgreen}{rgb}{0.0, 0.5, 0.2}
\definecolor{magenta}{rgb}{1,0,1}
\definecolor{violet}{rgb}{0.56,0,1}

\begin{document}

\title{\GRAthena: Binary Neutron Star Merger Simulations with Neutrino Transport}

\author{Boris \surname{Daszuta}$^1$\orcid{0000-0001-6091-2827}}
\email{boris.daszuta@uni-jena.de}
\author{Sebastiano \surname{Bernuzzi}$^1$\orcid{0000-0002-2334-0935}}
\email{sebastiano.bernuzzi@uni-jena.de}
\author{Maximilian \surname{Jacobi}$^1$\orcid{0000-0001-8168-4579}}
\author{Eduardo M. \surname{Guti\'errez}$^{2,3}$\orcid{0000-0001-7941-801X}}
\author{Peter \surname{Hammond}$^{5,4,2,3}$\orcid{0000-0002-9447-1043}}
\author{William \surname{Cook}$^1$\orcid{0000-0003-2244-3462}}
\author{David \surname{Radice}$^{2,3,6}$\orcid{0000-0001-6982-1008}}

\affiliation{${}^1$Theoretisch-Physikalisches Institut, Friedrich-Schiller-Universit{\"a}t Jena, 07743, Jena, Germany\\
  $^2$Institute for Gravitation and the Cosmos, The Pennsylvania State University, University Park, Pennsylvania, 16802, USA \\
  $^3$Department of Physics, The Pennsylvania State University, University Park, Pennsylvania, 16802, USA \\
  $^4$Department of Physics and Astronomy, University of New Hampshire, Durham, NH 03824, USA\\
  $^5$Max Planck Institute for Gravitational Physics (Albert Einstein Institute), 14476 Potsdam, Germany\\
  $^6$Department of Astronomy \& Astrophysics, The Pennsylvania State University, University Park, Pennsylvania, 16802, USA
}

\date{\today}

\begin{abstract}
We present general-relativistic radiation magnetohydrodynamics simulations of binary neutron star mergers performed with \GRAthena. Neutrino transport is treated using a moment-based, energy-integrated scheme (M1), augmented by neutrino number density evolution (N0). Our implementation is validated through an extensive suite of standard tests and demonstrated to perform robustly under adaptive mesh refinement.
As a first application, we simulate the gravitational collapse of a uniformly rotating, magnetized neutron star, demonstrating stable radiation evolution through apparent-horizon formation using a novel excision technique based on the tapering of state vector evolution inside the horizon.
To further test robustness in highly dynamic environments, we apply our code to two demanding binary neutron star merger scenarios. We investigate a long-lived remnant with the DD2 equation of state, evolved with full general-relativistic magnetohydrodynamics and M1 neutrino transport. Following this, a gravitational collapse scenario with the SFHo equation of state is explored. We showcase long-term stable evolution on neutrino cooling time-scales, demonstrating robust handling of excision and stable evolution of the post-collapse accretion phase in three-dimensional mergers with magnetic fields and neutrino radiation.

\end{abstract}

\pacs{
  04.25.D-,     
  04.30.Db,   
  95.30.Sf,     
  95.30.Lz,   
  97.60.Jd      
  97.60.Lf    
}

\maketitle

\section{Introduction}
\label{sec:intro}

The detection of gravitational waves from the binary neutron star merger GW170817~\cite{TheLIGOScientific:2017qsa}, together with its electromagnetic counterparts across the spectrum (a short gamma-ray burst~\cite{Monitor:2017mdv,GBM:2017lvd} and a kilonova powered by the radioactive decay of $r$-process nuclei~\cite{Abbott:2017wuw,Kasen:2017sxr}) has established binary neutron star (BNS) mergers as multi-messenger laboratories for dense-matter physics, heavy-element nucleosynthesis~\cite{Lattimer:1974slx}, and strong-field gravity.
The composition of the ejected material, and hence the observable kilonova signal~\cite{Metzger:2010sy,Metzger:2019zeh}, is strongly influenced by the electron fraction, which is in turn coupled to the neutrino irradiation of expanding ejecta. Accurate neutrino transport in numerical simulations of BNS mergers is therefore essential for connecting gravitational-wave observations to their electromagnetic counterparts and to predictions of $r$-process yields.

The neutrino radiation field in a BNS merger is, in principle, described by the general-relativistic Boltzmann equation. Direct solution in six-dimensional phase space in tandem with evolution of a spacetime remains prohibitively expensive. This has motivated a hierarchy of approximate treatments. 

At a first approximation, leakage schemes estimate local neutrino energy and (lepton) number emission rates without directly evolving the radiation field~\cite{Ruffert:1996by,Rosswog:2003rv}, interpolating between free emission and diffusion via an optical-depth estimate to capture bulk cooling. Extensions include improved optical-depth treatments, energy-dependent spectral leakage, radial-ray absorption, and general-relativistic and hybrid moment--leakage implementations~\cite{Neilsen:2014hha,Perego:2015agy,Ardevol-Pulpillo:2018btx,Sekiguchi:2010fh,Radice:2016dwd,Radice:2018pdn}. 

Moment-based methods instead
evolve angular moments of the (neutrino) distribution function entering the Boltzmann equation: the truncated two-moment (M1) formalism~\cite{Thorne:1981} can be cast into a form resembling the hydrodynamical equations~\cite{Shibata:2011kx}, particularly well-suited for numerical relativity. The (neutrino) energy density and flux are evolved, with closure of the moment scheme achieved through specification of a prescription for the pressure tensor.

Variants of moment-based neutrino transport have been implemented in a number of general-relativistic codes. The first application to BNS mergers was carried out with \textsc{SACRA}~\cite{Sekiguchi:2015dma} using a leakage plus transfer approach, while the \textsc{SpEC} code adopted an M1 scheme with an approximate treatment of the neutrino--matter coupling~\cite{Foucart:2015vpa}. In subsequent work~\cite{Foucart:2016rxm}, the evolution of the neutrino number density (the N0 extension of M1) was introduced, providing point-wise estimates of the average neutrino energy.
A semi-implicit scheme for the radiation transport, without ad-hoc approximation of the (stiff) source-terms, or requiring a fully implicit global solve was proposed by some of us~\cite{Radice:2021jtw}. This approach has since been adopted and extended by several groups~\cite{Foucart:2015vpa,Izquierdo:2022eaz,Schianchi:2023uky,musolino2024practicalguidemoment}.
Energy-dependent (multigroup) variants of M1 have been developed for core-collapse supernovae~\cite{Kuroda:2015bta,Roberts:2016lzn,Just:2015fda}, and more recently for post-merger remnants~\cite{cheong2023generalrelativisticradiationtransport,Cheong:2024buu}. This latter discretizes the neutrino energy spectrum into bins, capturing spectral-dependent opacities and improving predictions of ejecta composition at the cost of a proportional increase in the number of evolved fields~\cite{Cheong:2024buu}. Independent M1 implementations have since appeared within the
\textsc{THC}~\cite{Radice:2021jtw},
\textsc{MHDuet}~\cite{Izquierdo:2022eaz},
\textsc{BAM}~\cite{Schianchi:2023uky,supp:Neuweiler:2025klw},
\textsc{Gmunu}~\cite{cheong2023generalrelativisticradiationtransport},
and \textsc{FIL-M1}~\cite{musolino2024practicalguidemoment} frameworks.
Important differences among these implementations include the treatment of velocity-dependent source terms, choice of time-integration strategy, and the procedure for computing the variable Eddington factor from the closure relation.

In principle, truncated-moment schemes converge to the Boltzmann solution only in the limit of an infinite number of evolved moments. However based on direct comparison with Monte Carlo (MC) based solutions of the Boltzmann equation agreement is found at the $10\%$--$30\%$ level for global quantities such as neutrino luminosities, mean energies, and outflow electron fractions~\cite{Foucart:2024npn,Foucart:2020qjb}. Indeed the recent guided moments formalism~\cite{izquierdo2024guidedmomentsformalism} seeks a hybrid strategy by replacing the analytic M1 closure in optically thin regions with one computed from concurrent MC sampling, preserving moment-scheme accuracy in optically thick zones while alleviating the crossing-beam artifact of M1 in semi-transparent and free-streaming regimes~\cite{Foucart:2017mbt}.

We note, however, that incorporating additional physical mechanisms such as neutrino flavour transformation may substantially improve modelling fidelity beyond the aforementioned differences: the recent work of~\cite{Qiu:2025kgy} reports, in select configurations,  up to five-fold increases in neutron-rich ejecta when flavour oscillations are included.

These methodological advances have enabled a number of significant simulations.
Long-duration $\mathcal{O}(1)\,[\mathrm{s}]$ general-relativistic magnetohydrodynamics (GRMHD) evolution of BNS merger with approximate neutrino treatment has revealed the importance of magnetic-field amplification for jet launching and long-term remnant evolution~\cite{Kiuchi:2022nin}. In \cite{Combi:2022nhg}, GRMHD merger simulations were performed with a leakage-based weak-interaction treatment, characterizing $r$-process yields and electromagnetic signatures of the dynamical ejecta.
In~\cite{Musolino:2024sju}, GRMHD binary merger with full M1 neutrino transport was studied, demonstrating that neutrino heating modifies the onset of breakout. Recent grey-versus-spectral comparisons~\cite{Cheong:2024buu} have quantified the systematic biases inherent in energy-integrated schemes. Early fully general-relativistic BNS merger simulations with M1 transport~\cite{Sekiguchi:2015dma} showed that neutrino irradiation broadens the ejecta electron fraction. 
Subsequent studies of BNS and neutron-star--black-hole mergers~\cite{Foucart:2015vpa,Foucart:2015gaa,Foucart:2016rxm,Vincent:2019kor} refined ejecta predictions and highlighted the role of neutrino energy estimates. Longer M1 evolutions demonstrated improved modeling of neutrino-driven winds and diffusion relative to leakage schemes~\cite{Radice:2021jtw,Radice:2023zlw}, while axisymmetric and 3D simulations further explored neutrino-driven mass ejection~\cite{Fujibayashi:2017xsz,Fujibayashi:2020dvr}. 
More recently, Monte Carlo transport has been applied to collapsing BNS mergers, providing an independent benchmark for moment-based approaches~\cite{foucart2023generalrelativisticsimulations}. For a comprehensive review of neutrino transport methods in merger simulations we refer to~\cite{Foucart:2023liv}.%

Despite its successes, the energy-integrated (grey) M1 scheme has known limitations.
A typically adopted algebraic closure enforces a single propagation direction in the free-streaming limit, so intersecting beams produce unphysical radiation shocks and can overestimate polar neutrino densities~\cite{Foucart:2017mbt,Foucart:2018gis}. Moreover, grey schemes assume a neutrino spectrum to compute energy-averaged opacities; since $\kappa_a\propto \varepsilon_\nu^2$, hot free-streaming neutrinos in the ejecta can lead to strongly underestimated absorption unless the average energy is evolved (e.g., N0 extension)~\cite{Foucart:2016rxm,Foucart:2023liv}.
In spite of this, M1 transport may be a subdominant uncertainty relative to magnetic-field amplification, nuclear-physics inputs, and neglected effects such as flavour oscillations \cite{Qiu:2025kgy} or inelastic scattering~\cite{Foucart:2023liv}.

In this work, we present the implementation, validation, and first astrophysical applications of a grey M1+N0 neutrino-transport scheme within the adaptive-mesh-refinement (AMR) GRMHD code
\GRAthena{}~\cite{%
Daszuta:2021ecf,%
Cook:2023bag,%
Daszuta:2024ucu},
that closely follows the formulation of~\cite{Radice:2021jtw}.

Beyond the baseline M1+N0 scheme, we emphasise three key numerical (algorithmic) developments: (i) a low-order flux-correction (LOFC) strategy, inspired by \texttt{AthenaK} \cite{Stone:2024,Fields:2024pob} has been implemented in \GRAthena{};
(ii) conservative AMR with level-to-level flux corrections is applied to the M1+N0 sector;
(iii) a tapering-based excision procedure inside the apparent horizon (AH) is introduced, which allows stable evolution of both matter and radiation through and beyond gravitational collapse. The structure of the paper is as follows: we recall details of the M1+N0 formulation in \S\ref{sec:formulation}, followed by our numerical approach \S\ref{sec:num_method}. The core scheme is then validated against an extensive suite of test problems in \S\ref{sec:num_test}, including a cross-code comparison with \THC{}~\cite{Radice:2021jtw}. In \S\ref{sec:applications} we explore applications. Requisite diagnostics are first introduced (\S\ref{sbsec:diagnostics}), and in \S\ref{sbsec:rot_collapse} we study the gravitational collapse of a rotating magnetized neutron star. Subsequently we consider binary neutron star mergers where a scenario involving a long-lived remnant, and another that instead leads to rapid collapse post-merger are explored. This allows us to also investigate the effect of varying the Riemann solver (LLF vs HLLE) and magnetohydrodynamic treatment (HD vs MHD) (\S\ref{ssec:bns_calib},\,\S\ref{ssec:bns_prod}).
In \S\ref{sec:conclusion} we conclude.

\section{Formulation}
\label{sec:formulation}

Our goal in this section is to briefly summarise the M1+N0 scheme we have implemented that closely follows the approach of \cite{Radice:2021jtw} (see also related work \cite{Foucart:2015vpa,Foucart:2016rxm,Izquierdo:2022eaz,Schianchi:2023uky,musolino2024practicalguidemoment}), and to connect this to the GR(M)HD treatment of \GRAthena{} \cite{Daszuta:2021ecf,Cook:2023bag,Daszuta:2024chz}.

To fix nomenclature and the setting, we first recall our overall GR(M)HD treatment. We suppose that we have a space-time manifold endowed with metric and compatible connection $(\mathcal{M},\,g_{ab},\,\nabla{}_c)$ that is foliated by a family of non-intersecting, spatial hypersurfaces with members $\Sigma_t$ possessing the future-directed unit normal $n{}^a$. We utilize the standard $3+1$ ADM decomposition \cite{Arnowitt:1959ah}, where we denote\footnote{Ambient quantities take indices $a,\,b,\,\dots$ whereas projected quantities take $i,\,j,\,\dots$.}: induced metric on $\Sigma_t$ by $\gamma_{ij}$, the extrinsic curvature of $\Sigma_t$ by $K{}_{ij}$, and lapse and shift by $\alpha$ and $\beta{}^i$, respectively.

Our approach for treating dynamical space-time makes use of the Z4c formulation of numerical relativity \cite{Bernuzzi:2009ex,Hilditch:2012fp} with geometric fields\footnote{The definition and meaning of these variables is standard and may be found, together with our implementation details in \cite{Daszuta:2021ecf}.} $\big\{%
\alpha,\,\beta{}^i,\,
\chi,\,\tilde{\gamma}{}_{ij},\,\hat{K},\tilde{A}{}_{ij},\,\Theta,\,\tilde{\Gamma}{}^i%
\big\}$ evolved under cell-centered AMR as described in \cite{Daszuta:2024ucu}. We utilize the hydrodynamical formulation of \cite{Banyuls:1997zz}, where evolution is performed based on finite-volume (FV), flux-conservative, high-resolution shock-capturing (HRSC) techniques \cite{Thierfelder:2011yi}. Specifically, as detailed in \cite{Cook:2023bag}, we evolve the ($\sqrt{\gamma}$-densitized) quantities $\{\widetilde{D},\,\widetilde{S}{}_j,\,\widetilde{\tau},\,\widetilde{D Y}{}_{e}\}$, which are the baryonic matter density, momentum, energy density, and (densitized)-electron fraction respectively. This allows us to exploit finite-temperature, tabulated equation of state, as supplied by the \texttt{CompOSE} database \cite{Typel:2013rza}. When the fluid is magnetized, we assume ideal-MHD which we treat by leveraging an extension (to dynamical space-time) of the constrained-transport infrastructure of \Athena{} \cite{white2016extensionathenacode,%
felker2018fourthorderaccuratefinite,%
stone2020athenamathplusmathplus}. In contrast to our prior work, we have extended \GRAthena{} to utilize low-order flux-correction (LOFC), further supplemented by an approximate discrete-maximum-principle, motivated by the improved conservation properties observed in \cite{Stone:2024,Fields:2024pob}. Throughout this work our (M)HD treatment utilizes WENO5Z reconstruction \cite{Borges:2008a}, with interface fluxes combined according to the Local-Lax-Friedrichs (LLF) \cite{zanna2002efficientshockcapturingcentraltype} or HLLE \cite{doi:10.1137/1025002,doi:10.1137/0725021} prescription, with fall-back to PLM-van Leer \cite{vanleer1979ultimateconservativedifference}, and recomputation of LLF fluxes, in cells flagged for LOFC.

In this work we extend our infrastructure so as to additionally solve the energy-integrated Boltzmann equation, following the truncated moment formalism developed in \cite{Thorne:1981,Shibata:2011kx}. That results in a grey scheme. This we suitably augment, based on \cite{Foucart:2016rxm},  with an evolution equation for the neutrino number densities to recover point-wise information on average energies. Following \cite{Izquierdo:2022eaz}, we refer to this approach as ``M1+N0''.

\subsection{M1+N0 preliminaries}
\label{sbsec:prelim_bal_law}

To describe the neutrino species $\nu \in \{\nu_e,\,\overline{\nu}_e,\,\nu_x\}$ within the truncated-moment formalism \cite{Thorne:1981,Shibata:2011kx}, where $\nu_e$ and $\overline{\nu}_e$ denote electron neutrinos and anti-electron neutrinos, respectively, and $\nu_x$ collectively represents the heavy-lepton neutrinos and their antiparticles, an energy-integrated, (neutrino) radiation stress energy tensor ${}^{\mathrm{R}}T{}_{ab}$ is introduced.
It proves useful to supplement the Eulerian frame through the introduction of a fiducial\footnote{The truncated moment formalism is flexible in what may be chosen for this.
} four-velocity $u{}^a$. This allows us to decompose in Eulerian, and fiducial frames, for each $\nu$ (index suppressed) as
\begin{equation}
\begin{aligned}
  {}^{\mathrm{R}}\widetilde{T}{}_{ab}=\sqrt{\gamma}\,{}^{\mathrm{R}}T{}_{ab}
    &=
    \widetilde{E}\, n{}_a n{}_b + 2\widetilde{F}{}_{(a} n{}_{b)} + \widetilde{P}{}_{ab},\\
    &=
    \widetilde{J}\, u{}_a u{}_b + 2\widetilde{H}{}_{(a} u{}_{b)} + \widetilde{Q}{}_{ab};
\end{aligned}
\end{equation}
where elements of the triplets $(\widetilde{E},\,\widetilde{F}{}_a,\,\widetilde{P}{}_{ab})$ and $(\widetilde{J},\,\widetilde{H}{}_a,\,\widetilde{Q}{}_{ab})$ are $\sqrt{\gamma}$-densitized, and constitute the energy density, energy flux and pressure tensor in the respective frames. The projector orthogonal to $n{}^a$ is given by ${}^{\mathrm{E}}\mathcal{P}{}^a_b:=g{}^a{}_b+n{}^a n{}_b$ whereas for $u{}^a$ we have ${}^{\mathrm{F}}\mathcal{P}{}^a_b:=g{}^a{}_b+u{}^a u{}_b$. By construction, we have $n{}^a\perp\{\widetilde{F}{}_a,\,\widetilde{P}{}_{ab}\}$, i.e. $\widetilde{F}{}_a$ and $\widetilde{P}{}_{ab}$ are purely spatial and contravariant components satisfy $\widetilde{F}{}^0 = 0 = \widetilde{P}{}^{a0}$. Furthermore, if for some $X{}_a$ and $Y{} _a$ we have $n{}^a \perp \{X{}_a,\,Y{}_a\}$, then the Eulerian frame projector is idempotent, and $X{}^a Y{}_a = X{}^i Y{}_i$. In a similar vein $u{}^a\perp\{\widetilde{H}{}_a,\,\widetilde{Q}{}_{ab}\}$.

We fix now the choice of fiducial frame to be based on the fluid four-velocity: $u{}^a:=W(n{}^a+v{}^a)$, where $W:=-n{}^a u_a=\alpha u^0=1/\sqrt{1-v{}^iv{}_i}$ is the Lorentz factor and $v{}^a$ is the fluid spatial velocity (satisfying $n{}^a \perp v{}_a$). Furthermore, note that\footnote{In the language of \cite{Cook:2023bag}, ${}^E\mathcal{P}{}^i_a u{}^a=\tilde{u}{}^i$.} ${}^E\mathcal{P}{}^i_a u{}^a=u{}^i + W\beta{}^i/\alpha$.
These considerations lead to the following maps between frames:
\begin{equation}\label{eq:proj_fid2eul}
\begin{aligned}
  \widetilde{E}
  &=n{}_a n{}_b \widetilde{T}{}^{ab}
  =
  W^2\widetilde{J} +
  2 W v{}_a \widetilde{H}{}^a +
  v{}_a v{}_b \widetilde{Q}{}^{ab},\\
  \widetilde{F}{}_a
  &=-n{}_c{}^{ \mathrm E}\mathcal{P}{}_{ab} \widetilde{T}{}^{bc},\\
  &=
  W^2 v{}_a \widetilde{J} +
  (
    g{}_{ab} - n{}_a v{}_b
  )\left(
    W\widetilde{H}{}^b+
    v{}_c \widetilde{Q}{}^{bc}
  \right) \\
  &\hphantom{=}+
  W v{}_a v{}_b \widetilde{H}{}^b,\\
  \widetilde{P}{}_{ab}
  &=
  {}^{\mathrm E}\mathcal{P}{}_{ac}
  {}^{\mathrm E}\mathcal{P}{}_{bd}
  \widetilde{T}{}^{cd},
  \\
  &=
  W^2 v{}_a v{}_b \widetilde{J} +
  W (g{}_{ac}-n{}_a v{}_c) v{}_b \widetilde{H}{}^c\\
  &\hphantom{=}
  +
  W (g{}_{cb}-n{}_b v{}_c) v{}_a \widetilde{H}{}^c\\
  &\hphantom{=}
  +
  (g{}_{ac}-n{}_a v{}_c)(g{}_{bd}-n{}_b v{}_d)
  \widetilde{Q}{}^{cd};
\end{aligned}
\end{equation}
and similarly:
\begin{equation}\label{eq:proj_eul2fid}
\begin{aligned}
  \widetilde{J}
  &=
  u_a u_b \widetilde{T}^{ab}
  =
  W^2\left(
    \widetilde{E} -
    2\widetilde{F}^a v_a +
    \widetilde{P}^{ab} v_a v_b
  \right),\\
  \widetilde{H}^a
  &=
  -u_c\, {}^{\mathrm F}\mathcal{P}^a_b\, \widetilde{T}^{bc},\\
  &=
  W\left[
    (\widetilde{E} - \widetilde{F}^b v_b)
    {}^{\mathrm F}\mathcal{P}^a_c n^c +
    {}^{\mathrm F}\mathcal{P}^a_b
    (\widetilde{F}^b - v_c \widetilde{P}^{bc})
  \right],\\
  \widetilde{Q}^{ab}
  &=
  {}^{\mathrm F}\mathcal{P}^a_c\,
  {}^{\mathrm F}\mathcal{P}^b_d\,
  \widetilde{T}^{cd}.
\end{aligned}
\end{equation}
Notice that in the limit $v{}_i\rightarrow 0$ the two frames coincide. For later convenience, we also project Eq.\eqref{eq:proj_fid2eul} to write:
\begin{equation}\label{eq:dec_radflux}
  \begin{aligned}
    \widetilde{H}{}_{\mathrm{n}}
    &:= -\widetilde{H}{}^a n{}_a
    = W\left(
      \widetilde{E} -\widetilde{J} - \widetilde{F}{}_a v{}^a
    \right)
    , \\
    \widetilde{H}{}_a
    &=
    {}^E\mathcal{P}^b_a\widetilde{H}_b
    = W\left(
      \widetilde{F}{}_a - \widetilde{P}{}_{ab}v{}^b - \widetilde{J} v{}_a
    \right);
  \end{aligned}
\end{equation}
where $\widetilde{H}{}_a=\widetilde{H}{}_{\mathrm{n}} n{}_a + {}^E \mathcal{P}{}^b_a\widetilde{H}{}_b$. Furthermore, note that $u{}^a\perp \widetilde{H}{}_a$ leads to the expression $\widetilde{H}_{\mathrm{n}}=\widetilde{H}{}_i v{}^i$.

The full energy-momentum-stress tensor is now assumed to feature (magnetized) perfect-fluid and radiation contributions:  $T{}{}^{ab}={}^{\mathrm{F}} T{}^{ab} + \sum_\nu {}^{\mathrm{R}} T{}_{(\nu)}^{ab}$. Conservation of $T{}^{ab}$ leads to:
\begin{equation}
\label{eq:ems_cons}
\begin{aligned}
  \nabla{}_b\left[{}^{\mathrm{R}} T{}^{ab}\right] &= {}^{\mathrm{R}} S{}^{a}, &
  \nabla{}_b\left[{}^{\mathrm{F}} T{}^{ab}\right] &= -\sum_{\nu} {}^{\mathrm{R}} S{}_{(\nu)}^{a};
\end{aligned}
\end{equation}
where ${}^{\mathrm{R}} S{}_{(\nu)}^{a}$ is a term representing interaction between neutrinos and the fluid. For each of the species we adopt the general form:
\begin{equation}\label{eq:srcs_EF}
  {}^{\mathrm{R}} \widetilde{S}{}^{b}
  := \sqrt{\gamma}\,{}^{\mathrm{R}} S{}^{b}
  = (\sqrt{\gamma}\eta - \kappa{}_{\mathrm{a}} \widetilde{J}) u{}^b -
  (\kappa{}_{\mathrm{a}}  + \kappa{}_{\mathrm{s}} ) \widetilde{H}{}^b,
\end{equation}
where $\eta$, $\kappa{}_{\mathrm{a}}$ and $\kappa{}_{\mathrm{s}}$ are the (energy-averaged) neutrino emissivity, absorption and scattering opacities respectively.

At this point we can write down evolution equations for $\{\widetilde{E},\,\widetilde{F}{}_j\}$; however, we momentarily defer this to \S\ref{sbsec:eom_eqn}. Instead, we first address an important caveat whereby working with energy-averaged (grey) schemes discards information about the neutrino energy spectrum. Additionally, total lepton number of the system should be preserved under weak reactions. To this end, we follow a phenomenological prescription of \cite{Foucart:2016rxm}. For each neutrino species a number current $N{}^a$ and (fluid frame) number density $n:=-N{}^a u{}_a$ is introduced. This leads to the continuity equation:
\begin{align}\label{eq:divN_srcN}
  \nabla{}_a[N{}^a] &=  {}^{\mathrm{N}} S, & &
   {}^{\mathrm{N}} \widetilde{S} :=
   \sqrt{\gamma}\,{}^{\mathrm{N}} S = \sqrt{\gamma} \eta{}^0 - \kappa{}^0_{\mathrm{a}} \widetilde{n};
\end{align}
where $\kappa{}^0_{\mathrm{a}}$, and $\eta{}^0$, are the neutrino number absorption and emission opacities respectively. If we assume that the neutrino number and energy flux are aligned then we can adopt the closure relation:
\begin{equation}
  \begin{aligned}
    \widetilde{N}{}^a &= \widetilde{n} f{}^a, &\quad
    f{}^a &= u{}^a + \widetilde{H}{}^a / \widetilde{J};
  \end{aligned}
\end{equation}
where (with Eq.\eqref{eq:dec_radflux})
\begin{equation}
  \begin{aligned}
    \Gamma:=&
    -n{}_a f{}^a = \alpha f{}^0 = \frac{W}{\widetilde{J}} \left(
      \widetilde{E} - \widetilde{F}{}_a v{}^a
    \right),\\
    f{}^i :=& {}^E\mathcal{P}{}^i_a f{}^a
    = W\left(v{}^i - \beta{}^i / \alpha\right) + {}^E\mathcal{P}{}^i_a \widetilde{H}{}^a / \widetilde{J}.
  \end{aligned}
\end{equation}
Defining $\widetilde{N}:=\widetilde{n}\Gamma$ allows us to (approximately) compute the average energy $\langle \varepsilon \rangle$ of a given neutrino species, point-wise, in the fluid frame:
\begin{equation}
  \langle \varepsilon \rangle = \frac{\widetilde{J}}{\widetilde{n}}
  =
  \frac{W}{\widetilde{N}{}}
  \left(
    \widetilde{E} - \widetilde{F}{}^i v{}_i
  \right).
\end{equation}

\subsection{M1+N0 equations of motion}
\label{sbsec:eom_eqn}

The M1+N0 system may be written in the form of a balance law in the Eulerian frame \cite{Shibata:2011kx}. To this end, set  $\widetilde{U}{}_A:=(\widetilde{N},\,\widetilde{E},\,\widetilde{F}{}_k)$, where $0\leq A \leq 4$:
\begin{align}
  \label{eq:m1n0_balance_law}
  \partial_t[\widetilde{U}{}_A]+\partial_i[\widetilde{\mathcal{F}}{}_A^i]
  &= \widetilde{\mathcal{G}}{}_A + \widetilde{\mathcal{S}}{}_A;
\end{align}
where the flux vector term is
\begin{equation}\label{eq:m1flxvec}
  \widetilde{\mathcal{F}}{}^i_A = \alpha\left(
     \frac{\widetilde{N}}{\Gamma} f{}^i,\,
    \gamma{}^{ij}\widetilde{F}{}_j-\frac{\beta{}^i}{\alpha}\widetilde{E},\,
    \gamma{}^{ij}\widetilde{P}{}_{jk} - \frac{\beta{}^i}{\alpha} \widetilde{F}{}_k
  \right);
\end{equation}
geometric sources are given by
\begin{equation}
\begin{aligned}
  \tilde{\mathcal{G}}{}_0 &:=0,\\
  \tilde{\mathcal{G}}{}_1 &:=\alpha\left(
    \widetilde{P}{}^{ik}K{}_{ik} - \widetilde{F}{}^i \partial_i[\log(\alpha)]
  \right),\\
  \tilde{\mathcal{G}}{}_{1+k} &:=
    \widetilde{F}{}_i \partial_k[\beta{}^i]
    - \widetilde{E} \partial_k[\alpha]
    + \frac{\alpha}{2}\widetilde{P}{}^{ij} \partial_k[\gamma{}_{ji}];
\end{aligned}
\end{equation}
and collisional terms satisfy\footnote{Contractions appearing in the collisional sources carry a space-time index $a$.}:
\begin{equation}
  \label{eq:coll_sources}
  \widetilde{\mathcal{S}}{}_A :=
  \alpha \left(
    {}^N\widetilde{S}, \,
    -{}^R\widetilde{S}{}^b n{}_b,\,
    {}^E\mathcal{P}{}_{kb}\,{}^R\widetilde{S}{}^b   \right),
\end{equation}
which with Eq.\eqref{eq:dec_radflux}, Eq.\eqref{eq:srcs_EF}, and Eq.\eqref{eq:divN_srcN}  may be expanded to  %
\begin{equation}\label{eq:srcs_expanded}
\begin{aligned}
  \tilde{\mathcal{S}}{}_0
  &:=
  \alpha\left(
    \sqrt{\gamma} \eta{}^0 - \kappa{}^0_{\mathrm{a}} \widetilde{n}
  \right),\\
  \tilde{\mathcal{S}}{}_1
  &:=
  \alpha\left(
    W\left[\sqrt{\gamma}\eta-\kappa{}_{\mathrm{a}} \widetilde{J}\right]
    -\kappa{}_{\mathrm{as}} \widetilde{H}{}_{\mathrm{n}}
  \right),\\
  \tilde{\mathcal{S}}{}_{1+k}
  &:=
  \alpha\left(
    W\left[\sqrt{\gamma}\eta-\kappa{}_{\mathrm{a}} \widetilde{J} \right]v{}_k
    -\kappa{}_{\mathrm{as}}\,{}^E\mathcal{P}{}^b_k \widetilde{H}{}_{b}
  \right);
\end{aligned}
\end{equation}
where $\kappa{}_{\mathrm{as}}:=\kappa{}_{\mathrm{a}}+\kappa{}_{\mathrm{s}}$. Observe that while the evolved state vector $\widetilde{U}{}_A$ is written with respect to the Eulerian frame, sources involve terms that are written with respect to the fluid frame. Furthermore, while we have expressions governing the dynamics for each of the neutrino species in an associated triplet $(\widetilde{N},\,\widetilde{E},\,\widetilde{F}{}_k)$, the system remains under-determined. To close the system $\widetilde{P}{}_{ij}$ must still be fixed.

\subsection{Closure specification}
\label{sbsec:closure_spec}

The key idea of M1 \cite{Radice:2021jtw} is to close the truncated moment hierarchy by imposing an approximate relation on the radiation pressure tensor in terms of lower order moments, i.e. prescribing $\widetilde{P}{}_{ij}:=f[\widetilde{E},\,\widetilde{F}]{}_{ij}$. This can be achieved by considering behaviour in the limiting cases of optically thick (\S\ref{sbsbsec:thick_lim}) and thin (\S\ref{sbsbsec:thin_lim}) regimes, and connecting them (\S\ref{sbsbsec:interp_lim}) through imposition of a closure relation (\S\ref{sbsbsec:min_cl}). When combined, this furnishes us with an approximate solution to the transport problem\footnote{For a discussion on some of the limitations of this approach, we refer the reader to \cite{sadowski2013semiimplicitschemetreating,gavassino2020multifluidmodellingrelativistic}.}.

\subsubsection{Optically thick limit}
\label{sbsbsec:thick_lim}

The optically thick (diffusive) limit is characterised by radiation and matter being in thermodynamical equilibrium. In this section all quantities are taken to be in this regime. To proceed we impose isotropy on the radiation pressure tensor, which requires us to work in the fluid rest-frame:
\begin{equation}\label{eq:rad_fid_Q_iso}
  \widetilde{Q}{}_{ab} := \frac{1}{3} \widetilde{J}
  \left(
    g{}_{ab} + u{}_a u{}_b
  \right).
\end{equation}
With this prescription for $\widetilde{Q}{}_{ab}$, together with Eq.\eqref{eq:proj_fid2eul} and Eq.\eqref{eq:dec_radflux}, we find
\begin{equation}\label{eq:thick_fid2eul}
\begin{aligned}
  \widetilde{E}
  &=
  \frac{1}{3} \left( 4W^2 -1\right)\widetilde{J} + 2 W \widetilde{H}{}_{\mathrm{n}}
    ,\\
  \widetilde{F}{}_a
  &=
  \frac{4}{3} W^2 \widetilde{J} v{}_a +
  W\left(
    \widetilde{H}{}_{\mathrm{n}} v{}_a + {}^E \mathcal{P}{}_{ab} \widetilde{H}^b
  \right),\\
  \widetilde{P}{}_{ab}
  &=
  \frac{1}{3} \widetilde{J}\left(
    4 W^2 v{}_a v{}_b + {}^E \mathcal{P}{}_{ab}
  \right)
  + 2 W v{}_{(a}\,{}^E\mathcal{P}{}^c_{b)}\widetilde{H}{}_c.
\end{aligned}
\end{equation}
This completely determines Eulerian frame quantities in terms of their fluid frame counterparts. If we suppose that $H{}^a=0$ while working in the thick-limit, the last term in each expression of Eq.\eqref{eq:thick_fid2eul} vanishes. However, observe that $H{}^a=0$ does not necessarily entail that $\widetilde{F}{}_a$ is small in this regime \cite{Shibata:2011kx}.

To arrive at the complementary frame map we form the projection $u{}^a \widetilde{F}{}_a$, which together with Eq.\eqref{eq:dec_radflux} and Eq.\eqref{eq:thick_fid2eul} leads to
\begin{equation*}
\begin{aligned}
  \widetilde{J}
  &=
  \frac{3}{2W^2+1}\left(
    (2W^2 -1) \widetilde{E} - 2 W^2 \widetilde{F}{}_a v{}^a
  \right),\\
  \widetilde{H}{}_{\mathrm{n}}
  &=
  \frac{W}{2W^2+1}\left(
    4(1-W^2)\widetilde{E} + (4W^2-1)\widetilde{F}{}_a v{}^a
  \right),\\
  {}^E \mathcal{P}{}_{a}^b\widetilde{H}{}_b
  &=
  \frac{1}{W}\widetilde{F}{}_a -
  \frac{Wv{}_a}{2W^2+1}
  \left(
  4W^2 \widetilde{E} - (4W^2+1) \widetilde{F}{}_b v{}^b
  \right).
\end{aligned}
\end{equation*}
This allows us to assemble (as intermediate quantities) $\widetilde{J}=\widetilde{J}[\widetilde{E},\,\widetilde{F}]$, $\widetilde{H}{}^a=\widetilde{H}{}^a[\widetilde{E},\,\widetilde{F}]$, and in turn view $\widetilde{P}{}_{ab}$ as a functional of Eulerian frame quantities. 
\subsubsection{Optically thin limit}
\label{sbsbsec:thin_lim}

The optically thin (free-streaming) limit is characterised by zero optical depth; emissivity, absorption, and scattering are negligible. Consequently in this regime the collisional terms (Eq.\eqref{eq:coll_sources}) reduce to $\widetilde{\mathcal{S}}{}_A=0$. We assume that radiation propagates at the speed of light in the direction of the radiation flux:
\begin{equation}\label{eq:Pthin}
\begin{aligned}
  \widetilde{P}{}_{ab}
  &=
  \widetilde{E}  \widehat{F}{}_a \widehat{F}{}_b, &\quad
  \widehat{F}{}^a
  &:= \widetilde{F}{}^a / \Vert \widetilde{F} \Vert{}_\gamma;
\end{aligned}
\end{equation}
where for spatial quantities the norm is induced by the spatial metric $\gamma{}_{ij}$ via $\Vert \widetilde{F} \Vert^2_{\gamma} := \gamma{}_{ij} \widetilde{F}{}^i \widetilde{F}{}^j$. This limit is not unique, and care must be taken for compatibility with the underlying characteristics which leads to the choice of Eq.~\eqref{eq:Pthin} \cite{Shibata:2011kx}. This unfortunately can still result in pathologies (such as ``radiation shocks'' \cite{Radice:2021jtw}) for certain propagation problems. The main issue here can be understood by considering a collimated, colliding two-beam setup. Rather than cross, a resultant single beam emerges, free-streaming in the direction of the average of incident beam momenta \cite{Foucart:2015vpa,weih2020twomomentschemegeneralrelativistic}.

\subsubsection{Interpolation between regimes}
\label{sbsbsec:interp_lim}
In order to describe a state between thick (\S\ref{sbsbsec:thick_lim}) and thin (\S\ref{sbsbsec:thin_lim}) limits we interpolate. To this end, introduce the depth parameters $d{}_{\mathrm{T}}\in[0,\,1]$, $d_{\mathrm t}:=1-d_{\mathrm T}$, and write:
\begin{equation}
  \widetilde{P}{}_{ab} :=
  d{}_{\mathrm{t}} \widetilde{P}{}^{\mathrm{t}}_{ab}
  + d_{\mathrm{T}} \widetilde{P}{}^{\mathrm{T}}_{ab},
\end{equation}
where $\widetilde{P}{}^{\mathrm{T}}_{ab}$  and $\widetilde{P}{}^{\mathrm{t}}_{ab}$ are the thick and thin limit radiation pressure tensors, respectively.

This ansatz immediately allows us to factorize radiation frame moments:
\begin{equation}\label{eq:factorframe}
\begin{aligned}
  \widetilde{J}
  &=
  \widetilde{J}{}^{\mathrm{c}}
   +d{}_{\mathrm{t}} \widetilde{J}{}^{\mathrm{t}}
   +d{}_{\mathrm{T}} \widetilde{J}{}^{\mathrm{T}}, \\
  \widetilde{H}{}_{\mathrm{n}}
  &=
  \widetilde{H}{}^{\mathrm{c}}_{\mathrm{n}}
  +d{}_{\mathrm{t}} \widetilde{H}{}_{\mathrm{n}}^{\mathrm{t}}
  +d{}_{\mathrm{T}} \widetilde{H}{}_{\mathrm{n}}^{\mathrm{T}}, \\
  {}^E\mathcal{P}{}^b_a\widetilde{H}{}_b
  &=
  \left(
    \widetilde{H}{}^{\mathrm{c}}_{v}
    + d{}_{\mathrm{t}} \widetilde{H}{}^{\mathrm{t}}_{v}
    + d{}_{\mathrm{T}} \widetilde{H}{}^{\mathrm{T}}_{v}
  \right) v{}_a \\
  &\hphantom{=}
  +
  \left(
    \widetilde{H}{}^{\mathrm{c}}_{\widetilde{F}}
    + d{}_{\mathrm{t}} \widetilde{H}{}^{\mathrm{t}}_{\widetilde{F}}
    + d{}_{\mathrm{T}} \widetilde{H}{}^{\mathrm{T}}_{\widetilde{F}}
  \right) \widetilde{F}{}_a;
\end{aligned}
\end{equation}
where for $\widetilde{J}$ we have coefficients:
\begin{equation*}
\begin{aligned}
  \widetilde{J}{}^{\mathrm{c}}
  &:=
  W^2 \left(\widetilde{E} - 2 \widetilde{F}{}_a v{}^a\right),\\
  \widetilde{J}{}^{\mathrm{t}}
  &:=
  W^2\widetilde{E}
  \frac{\left(\widetilde{F}{}_a v{}^a \right) \left(\widetilde{F}{}_b v{}^b\right)}{\widetilde{F}{}_c \widetilde{F}{}^c},\\
  \widetilde{J}{}^{\mathrm{T}}
  &:=
  \frac{W^2-1}{2W^2+1}\left(
    4W^2 \widetilde{F}{}_a v{}^a
    + (3-2W^2)\widetilde{E}
  \right)
  ;
\end{aligned}
\end{equation*}
whereas for $\widetilde{H}{}_{\mathrm{n}}$:
\begin{equation*}
\begin{aligned}
  \widetilde{H}{}^{\mathrm{c}}_{\mathrm{n}}
  &:=
  -W \left(
    \widetilde{J}{}^{\mathrm{c}}
    + \left[
      \widetilde{F}{}_a v{}^a - \widetilde{E}
    \right]
  \right),\\
  \widetilde{H}{}_{\mathrm{n}}^{\mathrm{t},\mathrm{T}}
  &:=
  -W \widetilde{J}{}^{\mathrm{t},\mathrm{T}};
\end{aligned}
\end{equation*}
and finally:
\begin{equation*}
\begin{aligned}
  \widetilde{H}{}^{\mathrm{c},\mathrm{t}}_{v}
  &:=
  -W \widetilde{J}{}^{\mathrm{c},\mathrm{t}},\\
  \widetilde{H}{}^{\mathrm{T}}_{v}
  &:=
  -W\Big(
    \widetilde{J}{}^{\mathrm{T}}
    +
    \frac{1}{2W^2+1}\big[
      (3-2W^2)\widetilde{E} \\
  &\hphantom{=}
  + (2W^2-1)\widetilde{F}{}_a v{}^a
  \big]\Big),\\
  \widetilde{H}{}^{\mathrm{c}}_{\widetilde{F}}
  &=W, \quad
  \widetilde{H}{}^{\mathrm{t}}_{\widetilde{F}}
  =-W\frac{\widetilde{E} \widetilde{F}{}_a v{}^a}{\widetilde{F}{}_b \widetilde{F}{}^b},
  \quad
  \widetilde{H}{}^{\mathrm{T}}_{\widetilde{F}}
  =-W v{}_a v{}^a.
\end{aligned}
\end{equation*}

Thus for a fixed value of $d_{\mathrm{T}}$ and $\{\widetilde{E},\,\widetilde{F}{}_a\}$ we can compute $\{\widetilde{J},\,\widetilde{H}{}_a\}$ and $\widetilde{P}{}_{ab}$. We next turn to how the particular value of $d_{\mathrm{T}}$ may be dynamically specified based on the values $\widetilde{U}{}_A$ takes as it evolves, which completes a recipe for determining the equations of motion for M1+N0 (\S\ref{sbsec:eom_eqn}).

\subsubsection{Algebraic closure}
\label{sbsbsec:min_cl}

For the depth parameter introduced in \S\ref{sbsbsec:interp_lim} we adopt:
\begin{equation}
  d_T(\chi):=\frac{3(1-\chi(\xi))}{2}, \quad \chi\in\left[\frac{1}{3},\,1\right];
\end{equation}
where $\chi(\xi)$ is a closure function that must be prescribed (often also referred to as the variable Eddington factor).

A broad variety of choices for $\chi(\xi)$ have been surveyed in \cite{murchikova2017analyticclosuresm1}, with Minerbo (maximum entropy) \cite{minerbo1979mentmaximumentropy} shown to perform well on a broad selection of test problems. We adopt it, and accordingly set:
\begin{equation}
  \chi(\xi)
  :=
  \frac{1}{3}+
  \frac{\xi^2}{15}\left(6-2\xi+6\xi^2\right), \quad \xi\in\left[0,\,1\right];
\end{equation}
where $\xi\rightarrow 0$ in the thick-limit and $\xi\rightarrow 1$ in the free-streaming regime. In order to write $\xi$ as a functional of lower order moments, we follow \cite{Shibata:2011kx}:
\begin{equation}
  \label{eq:cl_par_spec}
  \xi^2
  :=
  \frac{\widetilde{H}{}_a \widetilde{H}{}^a}{\widetilde{J}^2}
  =\frac{\widetilde{H}{}_i \widetilde{H}{}^i- \widetilde{H}{}_{\mathrm n}^2}{\widetilde{J}^2}.
\end{equation}

This completes (an implicit) specification of $\widetilde{P}{}_{ab}$ thereby closing the M1+N0 system. In practice, an iterative, numerical approach allows for assembly of all required quantities (see \S\ref{sbsbsec:um_min_cl}).

\subsection{M1+N0 and GRMHD coupling}
\label{sbsec:ev_coupling}

As our aim is to describe a self-gravitating, magnetized fluid with neutrinos, we need to specify how M1+N0 is coupled to the GR(M)HD sector at the level of the evolution equations describing the latter.

On account of conservation of $T{}^{ab}$ (see Eq.\eqref{eq:ems_cons}) with ${}^{\mathrm{R}} S{}^a_{(\nu)}$ of Eq.\eqref{eq:srcs_EF}, the hydrodynamical evolution equations\footnote{We use the Valencia formulation \cite{Banyuls:1997zz,Anton:2005gi}, see Eqs.(4-7) of \cite{Cook:2023bag}.} for $\widetilde{\tau}$ and $\widetilde{S}{}_j$ pick up the additional source terms $\widetilde{\mathcal{S}}{}_{1}$ and $\widetilde{\mathcal{S}}{}_{1+j}$ of Eq.\eqref{eq:srcs_expanded} respectively, for each of the neutrino species. When magnetic fields are present, the form of the induction equation, following from $\nabla{}_a[\star F{}^{ab}]=0$ remains unchanged.

The lepton number density satisfies:
\begin{equation}
  \label{eq:leptcovd}
  \nabla{}_a[\rho Y{}_e u{}^a]
  = m{}_b \left(
    {}^{\mathrm{N}} S_{(\overline{\nu}_e)}
    -{}^{\mathrm{N}} S_{(\nu_e)}
  \right),
\end{equation}
where $\rho$ is the rest-mass density; $Y{}_e=n{}_e/n{}_b=n{}_p/n{}_b$ is the electron fraction; $n{}_e$, $n{}_b$, and $n{}_p$ are the electron, baryon, and proton number densities, respectively; $m{}_b$ is the reference baryon mass; and ${}^{\mathrm{N}} S_{(\nu)}$ is defined in Eq.\eqref{eq:divN_srcN}. In light of Eq.\eqref{eq:leptcovd} the $3+1$ GR(M)HD evolution equation governing $\widetilde{D} Y{}_e$ (see \cite{Cook:2023bag}) is modified to include an additional source term $m_b\left(\widetilde{\mathcal{S}}_{0,(\overline{\nu}_e)} - \widetilde{\mathcal{S}}_{0,(\nu_e)}\right)$ with $\widetilde{\mathcal{S}}{}_{0}$ of Eq.\eqref{eq:srcs_expanded}.

The ADM sources that couple matter fields to the geometric sector evolution we collect as ${}^{\mathrm{A}} S:=({}^{\mathrm{A}}\rho,\,{}^{\mathrm{A}} S{}_i,\, {}^{\mathrm{A}} S{}_{ij})$. When M1+N0 is activated, additional terms are added as:
\begin{equation*}
  {}^{\mathrm{A}} S \leftarrow
  {}^{\mathrm{A}} S + \frac{1}{\sqrt{\gamma}}\sum_{\nu\in\{\nu_e,\,\overline{\nu}_e,\,\nu_x\}}
  \left(
    \widetilde{E}{}_{(\nu)},\,
    \widetilde{F}{}_{(\nu),i},\,
    \widetilde{P}{}_{(\nu),ij}
  \right).
\end{equation*}

\subsection{Transport rates and equilibrium treatment}
\label{sbsec:opacities}

The transport rates (opacities) appearing in the collisional sources (Eq.\eqref{eq:srcs_expanded}) represent energy-averaged quantities that must be prescribed. In this work, these weak reaction rates, together with an average-energy based correction strategy \cite{Foucart:2016rxm}, are taken to be as in \cite{Radice:2021jtw,Zappa:2022rpd} (see also \cite{Perego:2019adq}). Specifically, we have ported \texttt{weakrates} from \THC{} \cite{Radice:2021jtw}. For convenience, we reproduce the specific processes accounted for in Tab.\ref{tab:wrates}.
\begin{table}
\caption{Weak reaction processes implemented in \GRAthena{}. In addition to neutrino species $\nu$, nucleons are denoted $N \in\{n, p\}$, whereas $A$ denotes a nucleus.}
\label{tab:wrates}
\begin{center}
  \begin{tabular}{ll}
\hline\hline
Reaction & Ref. \\
\hline
$\nu_e + n \leftrightarrow p + e^-$           & \cite{Bruenn:1985en} \\
$\bar{\nu}_{e} + p \leftrightarrow n + e^+$   & \cite{Bruenn:1985en} \\
$e^+ + e^- \rightarrow \nu + \bar{\nu}$       & \cite{Ruffert:1995fs} \\
$\gamma+\gamma \rightarrow \nu + \bar{\nu}$ & \cite{Ruffert:1995fs} \\
$N + N \rightarrow \nu + \bar{\nu} + N  + N$  & \cite{Burrows:2004vq} \\
$\nu + N \rightarrow \nu + N$                 & \cite{Ruffert:1995fs} \\
$\nu + A \rightarrow \nu + A$                 & \cite{Shapiro:1983du} \\
\hline\hline
\end{tabular}
\end{center}
\end{table}

\section{Numerical method}
\label{sec:num_method}

In order to evolve the composite GR(M)HD+M1+N0 system global methods are available \cite{Izquierdo:2022eaz}. Here however we prioritise reduced computational costs that retain simplicity, while reducing the number of infrastructure changes that are required within \GRAthena{}. This motivates the use of a split-step method (see e.g.~\cite{holden2010splitting}). Suppose we wish to evolve forward in time by $\delta t$.  At its simplest\footnote{The formal order of accuracy can be increased through systematic composition of many ($\delta t$-rescaled) split sub-steps \cite{hairer2006geometric,holden2010splitting}.} this entails freezing the dynamics of one of the sub-systems (M1+N0) and evolving GR(M)HD with a suitably selected integrator (for simplicity fixed as the SSPRK$(3,3)$ method of \cite{gottlieb2009highorderstrong} throughout, though in principle the choice is quite general). This is followed by evolution of the complementary part of the problem, M1+N0 (integrator details below), while GR(M)HD is frozen. Sources of the respective sub-systems and state-vectors are suitably updated after each such split $\delta t$.

\subsection{M1+N0 system}
\label{sec:M1N0system}

The M1+N0 system (Eq.\eqref{eq:m1n0_balance_law}), as it is written in balance-law form, as in the case of the hydrodynamical equations, it is natural to attempt a similar method-of-lines (MOL) strategy that features FV based HRSC for handling the formation of shocks. The behaviour of the M1+N0 equations in the optically thick regime however poses some additional numerical challenges. In the thick limit, the collisional sources can be stiff. When combined with well-posedness considerations in the diffusive limit \cite{hiscock1985genericinstabilitiesfirstorder,andersson2011consistentfirstordermodel} (see also \cite{Radice:2021jtw}) the use of an asymptotic preserving approach \cite{hu2017chapter5asymptoticpreserving} is motivated.

\subsubsection{Treatment of fluxes}
\label{sbsbsec:sp_disc}

In order to arrive at a conservative formulation of Eq.\eqref{eq:m1n0_balance_law}, we follow \cite{Radice:2021jtw}. The overall procedure may be understood by focusing on a single variable $U$, and direction $x$. Consider the following conservative finite-difference approximation at the cell-center $x\ixdi{}$:
\begin{equation}
  \partial_t[\widetilde{U}\ixdi{}] = -\frac{1}{\delta x}
  \left(
    \widetilde{\mathcal{F}}\ixdiph{} - \widetilde{\mathcal{F}}\ixdimh{}
  \right),
\end{equation}
where the grid is assumed to be uniformly spaced $\delta x:=x\ixdip{1}-x\ixdi{}$, $\widetilde{U}\ixdi{}:=\widetilde{U}(x\ixdi{})$, and $\widetilde{\mathcal{F}}\ixdipmh{}$ are cell-interface fluxes. To prepare these fluxes, a flux-splitting and flux-reconstruction approach is employed \cite{LeVeque1992}. The overall idea leverages hybridization and proceeds as follows: A second-order accurate, non-diffusive flux $\widetilde{\mathcal{F}}{}^{\mathrm{HO}}$, suitable for smooth regions which is asymptotic preserving, is combined with a standard (local) Lax-Friedrichs flux featuring diffusive correction $\widetilde{\mathcal{F}}{}^{\mathrm{LO}}$ which is first-order and robust near discontinuities.

Specifically we set:
\begin{equation}
\label{eq:flx_hyb}
  \widetilde{\mathcal{F}}\ixdiph
  =
  \widetilde{\mathcal{F}}\ixdiph^{\mathrm{HO}}
  - \Phi\ixdiph{}
  \left(
    \widetilde{\mathcal{F}}\ixdiph^{\mathrm{HO}}
    -\widetilde{\mathcal{F}}\ixdiph^{\mathrm{LO}}
  \right),
\end{equation}
where $\Phi\in[0,\,1]$ is a suitably selected flux-limiter function (defined below). Denote the nearest-neighbour average at a cell-interface of a sampled quantity $X$ by $\langle X \rangle\ixdiph:=(X\ixdip{1} + X\ixdi)/2$. Fluxes are chosen as
\begin{equation}
\begin{aligned}
  \widetilde{\mathcal{F}}\ixdiph^{\mathrm{HO}}
  &=
  \left\langle
    \widetilde{\mathcal{F}}[\widetilde{U}]
  \right\rangle\ixdiph{}, \\
  \widetilde{\mathcal{F}}\ixdiph^{\mathrm{LO}}
  &=
  \widetilde{\mathcal{F}}\ixdiph^{\mathrm{HO}}
  - \frac{\lambda\ixdiph}{2}
  \left(
    \widetilde{U}\ixdip{1} - \widetilde{U}\ixdi
  \right);
\end{aligned}
\end{equation}
where $\lambda\ixdiph$ is selected as a maximum over (local) characteristic speeds of the system, or a suitable approximation thereof\footnote{The M1 system characteristics are known in thin and thick limits analytically \cite{Shibata:2011kx}. During numerical experiments we have not however encountered substantial differences in performance and hence, while implemented, we do not make use of them in this work.}. Here we utilize the maximum of the speed of light between neighbouring cells. To this end define $c{}^{\pm}:=\left| \alpha \sqrt{\gamma{}^{xx}}\pm\beta{}^x \right|$. We then set $\lambda\ixdiph=\max \{ c\ixdi{}^\pm,\, c\ixdip{1}^\pm \}$.

The choice of the flux-limiter is based on $\mathrm{minmod}$:
\begin{equation}
  \varphi\ixdiph
  =
  \max
  \left(
    0,
    \min
    \left(
      1,\,
      \frac{\Delta\ixdim{2}[\widetilde{U}]}{\Delta\ixdim{1}[\widetilde{U}]},\,
      \frac{\Delta\ixdip{1}[\widetilde{U}]}{\Delta\ixdim{1}[\widetilde{U}]}
    \right)
  \right),
\end{equation}
where $\Delta\ixdi{}[\widetilde{U}]:=\widetilde{U}\ixdip{1} - \widetilde{U}\ixdi{}$. The form of $\Phi\ixdiph{}$ appearing in Eq.\eqref{eq:flx_hyb} is further subject to the requirements that: that spurious grid oscillations are not induced (even-odd decoupling) \cite{tang2018phenomenonartificialodd}, and diffusive correction is suppressed at high optical depth. If we first set:
\begin{equation}
\Phi\ixdiph = \hat{A}\ixdiph
  \left[
    1 - \varphi\ixdiph
  \right],
\end{equation}
then these two additional requirements can be respectively satisfied by selecting $\hat{A}\ixdiph{}=1$ if
\begin{equation}
  \label{eq:even_odd_dec}
  \left(\Delta\ixdim{2}[\widetilde{U}] \Delta\ixdim{1}[\widetilde{U}] < 0\right) \wedge
  \left(\Delta\ixdim{1}[\widetilde{U}] \Delta\ixdi{}[\widetilde{U}] < 0\right),
\end{equation}
and otherwise $\hat{A}\ixdiph{}=A\ixdiph{}$ with
\begin{equation}
  A\ixdiph{} := \min\left(
      1,\,
      \frac{1}{\delta x  \langle \kappa{}_{\mathrm{as}} \rangle \ixdiph}
    \right).
\end{equation}
The quantity $\delta x\,\langle \kappa{}_{\mathrm{as}}\rangle \ixdiph{}$ is a measure of the local optical depth\footnote{Other choices are possible here: For example selecting instead $A\ixdiph=\min(1,\,\langle \xi \rangle\ixdiph)$ when using the Minerbo closure also allows us to pass the numerical tests presented in \S\ref{sec:num_test}.}. Observe that if we are in a smooth region, at high optical depth, where the condition of Eq.\eqref{eq:even_odd_dec} is not satisfied, then $\widetilde{\mathcal{F}}\ixdiph \rightarrow \widetilde{\mathcal{F}}\ixdiph^{\mathrm{HO}}$, as required.

\subsubsection{Time-evolution: collisional sources}
\label{sbsbsec:coll_src}

In the optically thin regime, an approach that leverages an MOL based explicit time-integration of the semi-discretized form of Eq.\eqref{eq:m1n0_balance_law}, with fluxes treated as described in \S\ref{sbsbsec:sp_disc}, does not pose particular difficulties. On the other hand, in the optically thick regime the absorption $(\kappa{}_{\mathrm{a}})$ and scattering $(\kappa{}_{\mathrm{s}})$ opacities, appearing in the sources (Eq.\eqref{eq:srcs_expanded}), can reach large values. Consequently these terms become ``stiff'' and to satisfy stability conditions an explicit time-integrator would require (prohibitively small) time-steps of the order $\mathcal{O}(1/\kappa)$ (see the discussion in e.g.~\cite{Izquierdo:2022eaz}).

To tackle this issue, we adopt the two-step, semi-implicit scheme \cite{Radice:2021jtw}. In contrast to many other approaches \cite{Foucart:2016rxm,weih2020twomomentschemegeneralrelativistic,Schianchi:2023uky}, we fully treat non-linear terms appearing in the radiation-matter coupling as in \cite{Radice:2021jtw,Izquierdo:2022eaz,musolino2024practicalguidemoment}. Consider a fixed grid-point $x\ixdi{}$. When opacities locally satisfy $\delta t > 1 / \kappa{}_{\mathrm{a}}$ or $\delta t > 1 / \kappa{}_{\mathrm{s}}$ then instead of explicit integration, we make use of the update rule:
\begin{equation*}
\begin{aligned}
  \frac{\widetilde{U}{}^* - \widetilde{U}{}^{(k)}}{\widehat{\delta t}_1}
  &=
  -\partial{}_i[\widetilde{\mathcal{F}}{}^i[\widetilde{U}{}^{(k)}]]
  +\widetilde{\mathcal{G}}[\widetilde{U}{}^{(k)}]
  +\widetilde{\mathcal{S}}[\widetilde{U}{}^{*}],\\
  \frac{\widetilde{U}{}^{(k+1)} - \widetilde{U}{}^{(k)}}{\widehat{\delta t}_2}
  &=
  -\partial{}_i[\widetilde{\mathcal{F}}{}^i[\widetilde{U}{}^{*}]]
  +\widetilde{\mathcal{G}}[\widetilde{U}{}^{*}]
  +\widetilde{\mathcal{S}}[\widetilde{U}{}^{(k+1)}];
\end{aligned}
\end{equation*}
where $\widehat{\delta t}_1:=\delta t / 2$ on the first sub-step and $\widehat{\delta t}_2:=\delta t$ on the second. On each sub-step we have a (non-linear) system of equations:
\begin{equation}\label{eq:nlsiupdatesystem}
\begin{aligned}
  \widetilde{\mathcal{Z}}{}_A[\widetilde{U}{}^*]
  &:=
  \widetilde{U}{}^*_A - \widehat{\delta t}_s \widetilde{\mathcal{S}}{}_A[\widetilde{U}{}^*]
  - \widetilde{W}{}_A, \\
  \widetilde{W}{}_A
  &:=
  \widetilde{U}{}_A + \widehat{\delta t}_s
  \left(
    -\partial{}_i\left[\widetilde{\mathcal{F}}{}^i_A\right]
    + \widetilde{\mathcal{G}}{}_A
  \right);
\end{aligned}
\end{equation}
where $\widetilde{W}{}_A$ is independent of the implicit term $\widetilde{U}{}^*_A$. To tackle Eq.\eqref{eq:nlsiupdatesystem} numerically, we form a Newton sequence of approximants:
\begin{equation*}
  {}^{[i+1]} \widetilde{U}{}^*_A
  \longleftarrow
  {}^{[i]} \widetilde{U}{}^*_A
  -\left(
    \widetilde{\mathcal{J}}^{-1}
  \right){}_A{}^B
  \widetilde{\mathcal{Z}}{}_B\left[{}^{[i]} \widetilde{U}{}^*\right],
\end{equation*}
where $\widetilde{\mathcal{J}}{}^B{}_A:=\delta \widetilde{\mathcal{Z}}{}_A / \delta \widetilde{U}{}^*_B=\delta{}^B_A-\widehat{\delta t}_s \left(\delta \widetilde{S}{}_A / \widetilde{U}{}^*_B\right)$ is the (formal) Jacobian (see Appendix~\S\ref{sec:appendix_jacobian}). During each step of the sequence, we also solve an embedded problem for the closure function (see \S\ref{sbsbsec:um_min_cl}). The iterative procedure is continued until $\Vert {}^{[i+1]} \widetilde{U}{}^* - {}^{[i]} \widetilde{U}{}^*  \Vert < \epsilon$ in suitable norm\footnote{To achieve this we utilize the \texttt{HybridsJ} method of \texttt{GSL} \cite{galassi2018gnu}.}. As a fall-back mechanism in the event that a solution fails to converge, we revert to direct linearisation of the equations in the thick limit. 
In order to improve efficiency of the solver, a judicious initial guess can be provided as in \cite{Radice:2021jtw}. In brief, from the state vector, we take Eulerian frame quantities and transform to the fluid frame. These are then approximately evolved as \cite{Mezzacappa:2020oyq}:
\begin{equation}
\begin{aligned}
  {}^{[0]}\widetilde{J}
  =&
  \widetilde{J} + \frac{\delta t}{W}(\sqrt{\gamma}\eta - \kappa{}_{\mathrm{a}} \, {}^{[0]}\widetilde{J})
  ,\\
  {}^{[0]}\widetilde{H}{}_j
  =&
  \widetilde{H}{}_j - \frac{\delta t}{W} \kappa{}_{\mathrm{as}}\,{}^{[0]}\widetilde{H}{}_j
  ;\\
\end{aligned}
\end{equation}
which can be rearranged for an initial guess in the fluid frame. By assuming the thick limit, we can prepare $\left\{{}^{[0]}\widetilde{E},\,{}^{[0]}\widetilde{F}{}_j\right\}$ based on Eq.\eqref{eq:thick_fid2eul}.

In contrast the component $\widetilde{\mathcal{Z}}_0$ of Eq.\eqref{eq:nlsiupdatesystem}, representing neutrino species, decouples and may be solved for directly. The first sub-step reads:
\begin{equation}
  \label{eq:neutrino_dens_evo}
  \widetilde{N}{}^*
  =
  \frac{\widetilde{N} + \widehat{\delta t}_1 \left(
    \alpha \sqrt{\gamma} \eta{}^0
    - \partial{}_i\left[
      \widetilde{\mathcal{F}}{}^i_0[\widetilde{N}]
    \right]
  \right) }{1 + \widehat{\delta t}_1 \alpha \kappa{}^0_{\mathrm{a}} \Gamma[\widetilde{E}^*,\,\widetilde{F}^*]^{-1}},
\end{equation}
with the second sub-step taking a similar form. Observe that the updated $\{\widetilde{E}^*,\,\widetilde{F}{}^*_j\}$ are utilized in Eq.\eqref{eq:neutrino_dens_evo}.

Finally, we remark that, in order to avoid issues associated with loss of significance during division, we impose (small but finite) floors\footnote{We fix $\mbox{\texttt{floor}}(\widetilde{E})=10^{-40}=\mbox{\texttt{floor}}(\widetilde{N})$.} on $\widetilde{E}$ and $\widetilde{N}$. Additionally, as explained in \cite{Schianchi:2023uky,musolino2024practicalguidemoment}, for robustness, we enforce a causality condition $\widetilde{E} < \Vert \widetilde{F} \Vert_\gamma$ via rescaling of $\widetilde{F}{}_j$. During the evolution, a candidate update for a state at the point $x\ixdi{}$ is constructed. If any of the aforementioned conditions must be imposed, neighbouring interface fluxes are reverted to $\widetilde{\mathcal{F}}{}^{\mathrm{LO}}$ by setting $\Phi\ixdiph{}=1$ in Eq.\eqref{eq:flx_hyb}. This may be viewed as a form of \textit{flux-correction}. The effect of this is demonstrated in \S\ref{sbsec:scattering_limit}.

\subsubsection{Time-evolution: source-limiting and equilibrium treatment}
\label{sbsbsec:src_eql}

In the context of binary neutron star mergers, two further ingredients beyond the semi-implicit scheme of \S\ref{sbsbsec:coll_src} have been found to improve stability.  First, following~\cite{Radice:2021jtw}, updates to the radiation state vector are limited. Thus if the semi-implicit solve would yield changes to the neutrino state vector that result in unphysical values, or lead to an electron fraction outside the bounds of the EOS table, the (source) update is first rescaled prior to its application to the state-vector. Second, in regions where the radiation-matter equilibration timescale $\tau := (\sqrt{\kappa_{\mathrm{a}}(\kappa_{\mathrm{a}}+\kappa_{\mathrm{s}})})^{-1}$ is much shorter than the time-step, that is, when $\delta t$ exceeds $X$ e-foldings of $\tau$, then the neutrino average energies entering the number density evolution are assumed to be at their equilibrium values~\cite{Zappa:2022rpd}.  Without this treatment, the solver can develop spurious features in high-density, low-$Y_{\mathrm{e}}$ regions during the first few milliseconds after merger~\cite{Zappa:2022rpd}.  The threshold parameter $X$ is problem-dependent and must be tuned according to the demands of a simulation, with typical values of $X=10$--$20$ selected in~\cite{Zappa:2022rpd}.  We use an alternative comparably effective prescription, where for fixed $X=1$ if cells are flagged at equilibrium then their nearest-neighbours are also, thus instead effectively smearing the equilibrium mask.

\subsubsection{Solution of parametrized algebraic closure}
\label{sbsbsec:um_min_cl}
In order to fix the value the closure function takes, we introduce a residual based on Eq.\eqref{eq:cl_par_spec}:
\begin{align}
  \mathcal{Z}_{\xi}
  :=
  \frac{1}{\widetilde{E}^2}\left[
    \xi^2 \widetilde{J}^2 - \widetilde{H}{}_a\widetilde{H}{}^a
  \right].
\end{align}
The following algorithm then allows us to generate a sequence of approximations ${}^{[i]}\chi$:
\begin{itemize}
  \item Compute $\widetilde{P}{}^{\mathrm{T}}_{ab}$  and $\widetilde{P}{}^{\mathrm{t}}_{ab}$, and based on the current ${}^{[i]}\chi$ assemble $\widetilde{P}{}_{ab}$.
  \item{Compute $\{\widetilde{J},\,\widetilde{H}{}_a\}$ based on Eq.\eqref{eq:proj_eul2fid} and Eq.\eqref{eq:dec_radflux}.
}
  \item Assemble $\widetilde{H}{}^a \widetilde{H}{}_a = \widetilde{H}{}^i \widetilde{H}{}_i-\widetilde{H}{}_{\mathbf{n}}^2$ and$\{\xi,\,\chi\}$.
  \item Check $|\mathcal{Z}_{\xi}|<\varepsilon_R$; if not, compute new guess for $\chi$ and go back to second step.
\end{itemize}
In our code, we employ a Newton root-finder. Derivatives of $\mathcal{Z}_{\xi}$ are required and may be assembled based on the expressions of \S\ref{sbsbsec:interp_lim}. If during the course of the iteration $\xi\notin[0,\,1]$, we fall-back to a derivative-free method, such as Illinois \cite{dowell1971illinois}, or Brent-Dekker \cite{galassi2018gnu}. 

\subsection{Excision strategy}
\label{sbsec:ahf_excision_tap}

To study the long-term properties of a merger remnant, that undergoes collapse, we require robust detection and handling of regions that develop an apparent horizon (AH). As is well-known, in such regions, matter fields can attain extreme values, that potentially pose challenges for utilization of tabulated EOS. We seek to rectify this issue by suitably excising such regions in a \textit{controlled} fashion. 
During the course of a simulation, the location $\mathbf{c}^*_A(t)$ $(A:=1,\,2)$ of local extremal (minimal) values of the lapse for each constituent of the binary are tracked. In the post-merger phase, the ``fast-flow'' AH detection algorithm of \cite{Gundlach:1997us} is employed on-line. Suppose an AH is formed, at time $t_{\mathrm{AH}}$. The trajectory $\mathbf{c}^*(t)$ then approximates the center of the AH at times $t\geq t_{\mathrm{AH}}$. The fast-flow algorithm allows us to additionally characterise AH quantities such as: minimal radius $r^*(t)$, gravitational mass, and dimensionless spin.

Denote the ball of minimal radius, with center $\mathbf{c}^*(t)$ associated with the AH by $\mathcal{B}_{r^*}(t;\,\mathbf{c}^*)$. On the interior of $\mathcal{B}_{r^*}$, i.e.~ within the AH, over a region of compact support, we modify the M1(+N0) (and hydrodynamical) equations of motion such that the state-vector $\widetilde{U}$ is driven towards atmosphere values $\widetilde{U}{}^{\mathrm{atm}}$. To this end, we introduce source terms $-\mathcal{A}(t,\,x) \left[\widetilde{U}-\widetilde{U}{}^{\mathrm{atm}}\right]$, in the (matter) equations of motion, where $\mathcal{A}$ is a tapering function. We have found it helpful to also follow a similar prescription for the M1+N0 sector. For simplicity, suppose $r$ denotes the distance from the center of the AH, then  we restrict ourselves to the following form of taper:
\begin{equation*}
\begin{aligned}
  \mathcal{A}(t,\,r) &:=
  \lambda
  \left(
    1-
    \left[
      1 - \exp\left(1 + \frac{1}{f(r)^2-1}\right)
    \right]^p
  \right),\\
  f(t,\,r) :=& \min\left(1-\varepsilon{}_{\mathrm{fl}},\,r / (f_r r^*(t))\right);
\end{aligned}
\end{equation*}
where strength may be controlled by the the parameters $\lambda$, $p$, and $f_r$; and $\varepsilon{}_{\mathrm{fl}}$ is a numerical floor. Suitably tuning the aforementioned control parameters helps mitigate development of extremal values in matter fields, that would encroach on the limits of the tabulated EOS employed in this work. 
Unless otherwise stated we use a typical choice of parameters $\lambda=100$, $p=10$, and $f_r=1/3$.

\section{Implementation validation}
\label{sec:num_test}

Having described the overall formulation, together with numerical approach, we now verify the robustness of our implementation within \GRAthena{} over a variety of test problems. The idea here is to focus on distinct regimes (e.g.~non-stiff vs stiff, shocked vs smooth) through a collection of now standard model problems (see \cite{Radice:2021jtw,Izquierdo:2022eaz,Schianchi:2023uky,musolino2024practicalguidemoment} and references therein).
Additionally, we perform cross-code comparison, isolating focus on our ported opacity library directly.

To set the stage for what follows: in \GRAthena{}, a target computational domain $\Omega$, the \Mesh{}, is selected, with overall spatial extent and number of samples along each spatial direction fixed. Subsequently, decomposition into sub-domains $\Omega_i$ (\MeshBlock{} objects), that exactly partition the \Mesh{}, is performed. Sampled field data is communicated between \MeshBlock{} objects through the introduction of a number of ghost-nodes\footnote{We fix $N_g=4$ throughout this work.} $N_g$.
When the grid is further (dynamically) refined\footnote{Along each axis a \MeshBlock{} is split to two daughter objects, at fixed sampling, but double the resolution. The procedure is recursive, and constrained to maintain a refinement ratio of (at-most) $2:1$.}, there will exist interfaces with a \MeshBlock{} either-side at differing levels of refinement. Here, level-to-level transfer of sampled M1(+N0) variable data, populating ghost-layers, follows the same strategy as the conservative, cell-centered hydrodynamical treatment, together with associated \MeshBlock{}-interface flux correction detailed in \cite{Stone:2020}.

\subsection{Optically thin regime: advection}
\label{sbsec:opt_thin_advection}

We start by considering propagation of a beam of radiation in an optically thin medium as in \cite{Radice:2021jtw}. The \Mesh{} is set to be a thin layer, three-dimensional Cartesian grid. We take $x\in[-4,\,4]$ with $N_M$ samples. In the $y$ and $z$ directions we sample with four points, and select overall extent such that $\delta x=\delta y=\delta z$. The time-step is chosen so as to satisfy a Courant-Friedrich-Lewy condition (CFL) of $0.5$.

Initial data (ID) is provided by the choice $\widetilde{E}\big|_{t=0}= \widetilde{F}^x\big|_{t=0} = H(x+1/2)$, where $H$ is the Heaviside (unit-step) function. The optically thin condition entails $\kappa{}_{\mathrm{s}}=\kappa{}_{\mathrm{a}}=\eta=0$. The background fluid velocity is taken to have a single non-zero component $v{}^x$. This latter, is specified piece-wise: $v{}^x=-0.87$ (for $x>0$), and $v{}^x=+0.87$ (for $x<0$). This induces a relative Lorentz factor of $W_d=7$ between the two parts of the domain, and a factor of $W_g=2$ in the grid frame. As our closure is prescribed in the fluid frame, this is a demanding configuration for the scheme. Outflow boundary conditions (BC) for the $x$ direction are set, and periodicity is assumed in $y$ and $z$. The background space-time is taken to be flat.

\begin{figure}[t]
  \centering
    \includegraphics[width=0.49\textwidth]{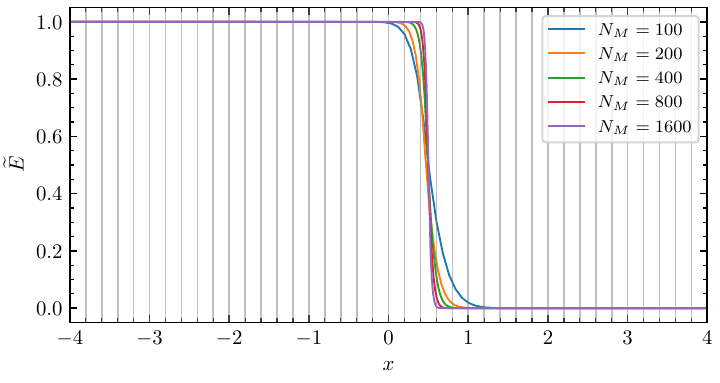}
    \caption{Optically thin advection test: Euler-frame energy density $\widetilde{E}$ at time $t=2$, for a variety of grid resolutions. Note the absence of any profile artifacting from the the fluid velocity interface (at $x=0$). The {\Mesh{}} sampling is indicated in the inset.
    For each run the {\MeshBlock{}} sampling (along the $x$-direction) is $N_B=10$, with boundaries demarcated in gray for the $N_M=400$ run.
    }
 \label{fig:opt_thin_advection}
\end{figure}

As can be seen in Fig.~\ref{fig:opt_thin_advection}, our implementation can correctly capture advection of the pulse through the velocity discontinuity at a variety of resolutions. To assess convergence, we evolve to a final time of $t=2$ with \Mesh{} sampling of $N_M=1600$ and fix this as a reference solution. Utilizing $\widetilde{E}$ ID profiles, that have been evolved with spatial sampling $N_M\in\{200,\,200,\,400,\,800\}$, and comparing at $t=2$, leads to an estimated convergence order, in the discrete $L^2$-norm, of ${\sim} 1.49$. This is compatible with the step-character of the ID and advected solution. Thus transport through a moving medium is handled, without inducing artificial oscillations due to discontinuity.

\subsection{Scattering dominated limit}
\label{sbsec:scattering_limit}

To model the behaviour of neutrinos diffusing from a BNS merger remnant, it is crucial to capture the asymptotic diffusion limit of the M1(+N0) equations. We consider two effectively one-dimensional problems, on a flat geometry, in a purely scattering medium: constant density $\rho=1$ (arbitrary units) is set, together with $\kappa{}_{\mathrm{s}}=10^3$ and all other opacities set to zero. For this subsection CFL is fixed at $0.65$.

\subsubsection{Diffusion of top-hat profile}
\label{sbsec:diff_lim}

For this problem (see \cite{Pons:2000br,McKinney:2013txa,Kuroda:2015bta,weih2020twomomentschemegeneralrelativistic,Radice:2021jtw,Izquierdo:2022eaz}), we again set the computational domain analogously to the advection problem. A sequence of runs of increasing resolution by selecting \Mesh{} sampling to be $N_M\in\{100,\,200,\,400,\,800,\,1600\}$ is considered here. For ID, we take $\widetilde{E}\big|_{t=0}=H(x+1/2)-H(x-1/2)$, and  $\widetilde{F}{}^i\big|_{t=0}=0=v{}^i\big|_{t=0}$.

Based on these choices it is clear that  we are in the optically thick regime as $\kappa{}_{\mathrm{s}} \delta x \gg 1$, and consequently, we are in a regime where mean-free-path is small compared with the grid spacing.

\begin{figure}[t]
  \centering
    \includegraphics[width=0.49\textwidth]{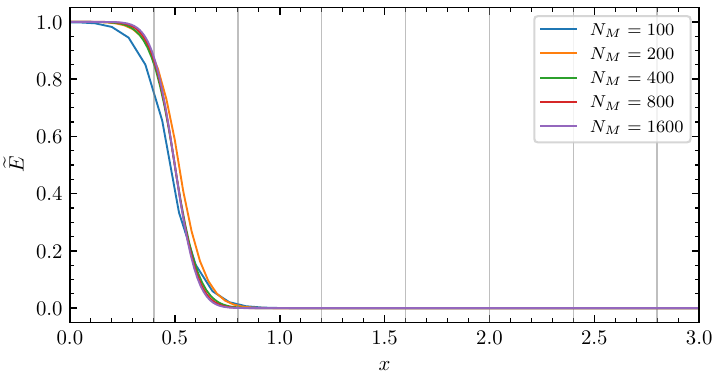}
    \caption{Diffusion test: Euler-frame energy density $\widetilde{E}$ at time $t=10$, for a variety of grid resolutions. Inset indicates differing number of {\Mesh{}} samples for distinct simulations. We have $x\in[-4,\,4]$ during calculation; for this plot, however, we suppress (symmetric) data for $x<0$ to better illustrate differences in the runs.
    For each run, the {\MeshBlock{}} sampling is $N_B=10$, with boundaries demarcated in gray for the $N_M=400$ run.
    }
 \label{fig:diffusion_test}
\end{figure}

As can be seen in Fig.\ref{fig:diffusion_test}, as resolution is increased, the numerical solutions approach that of the $N_M=1600$ run. Approximate analytical solutions of this problem are known in this regime, based on assuming that the evolution of $\widetilde{E}$ is well-described by a simple diffusion equation (see e.g~\cite{Izquierdo:2022eaz}). For simplicity, however, we consider self-convergence using a numerical reference profile of $\widetilde{E}$ as was done for the advection problem. Working with the discrete $L^2$-norm and profiles evolved to a final time $t=10$, leads to an estimate for the convergence factor of $\sim 2.09$. This is compatible with the overall design order of the scheme.

\subsubsection{Diffusion in a moving medium}
\label{sbsec:diff_mm_lim}

For this problem (see \cite{Nagakura:2014nta,chan2020novelmultidimensionalboltzmann,Radice:2021jtw,Izquierdo:2022eaz,Schianchi:2023uky}), in addition to the resolution sequences selected for the computational domain, we will exercise our adaptive-mesh-refinement (AMR) capabilities. Initial data is selected so as to describe a right-ward propagating Gaussian profile in $\widetilde{E}$ under the assumption that radiation is fully trapped, which is characterized by $\widetilde{H}{}^a=0$. Specifically, we set $\widetilde{E}\big|_{t=0}=\exp(-9x^2)$, together with $\widetilde{F}{}_i\big|_{t=0}=v{}_i / (1-(4w^2)^{-1}) \times \widetilde{E}\big|_{t=0}$, where we have used Eq.\eqref{eq:thick_fid2eul}. The non-zero component of velocity is taken to be $v{}^x=0.5$.

\begin{figure}[t]
  \centering
    \includegraphics[width=0.49\textwidth]{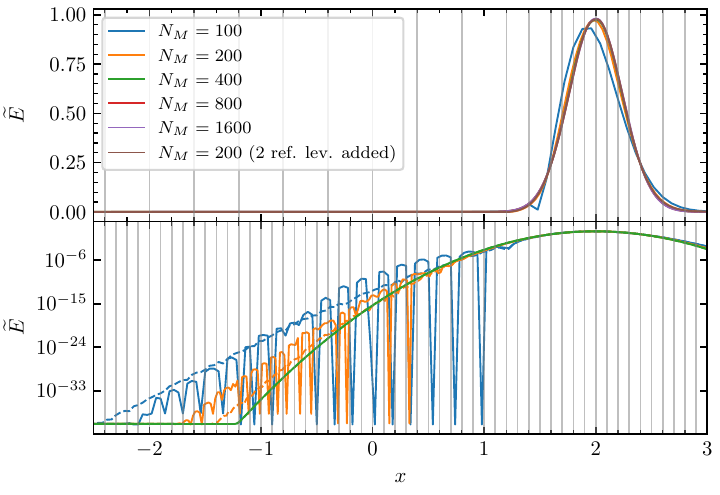}
    \caption{
      Diffusion test, moving-medium: Euler-frame energy density $\widetilde{E}$ at time $t=4$, at a variety of grid resolutions. (Upper) Observe clear convergent trend as number of {\Mesh{}} samples is increased. A run at $N_M=200$ featuring two additional levels of adaptive refinement, is also shown. The AMR criterion is dynamical, and traces the field local maximum. Absence of refinement boundary artifacts indicates level-to-level transfer of data, and refluxing are robust.
      (Lower) Selected runs in $\log$-scale comparing (de)-activation of FC of the M1-sector in (solid) dashed lines respectively. Observe absence of trailing-edge artifacting with FC. (Common) {\MeshBlock{} sampling is $N_B=10$} with boundaries demarcated in gray for the AMR run in the upper panel, and the $N_M=400$ run in the lower panel.
    }
 \label{fig:diffusion_test_mm_split}
\end{figure}

In Fig.\ref{fig:diffusion_test_mm_split}, we observe that the initial Gaussian profile, maintains approximately unit height in $\widetilde{E}$, while advecting and diffusing to its final state at $t=4$, centered at $x=2$. This agrees with the general behaviour reported in \cite{Radice:2021jtw,Izquierdo:2022eaz,Schianchi:2023uky,musolino2024practicalguidemoment}. Working with the discrete $L^2$-norm and uniform resolution profiles evolved to a final time $t=4$, leads to an estimate for the convergence factor of ${\sim} 2.50$. In order to provide an initial test of our AMR treatment of the M1 variables, we attach a tracker $\mathbf{c}(t)$ to the maximum of $\widetilde{E}(t)$, and impose that two levels of refinement are added for points satisfying $\Vert \mathbf{x}(t) - \mathbf{c}(t)\Vert \leq 0.25$. This is shown for a run with $N_M=200$ in Fig.\ref{fig:diffusion_test_mm_split} (upper). As can be seen there the additional levels capture the profile as it evolves without introduction of inter-level artifacts.

When a candidate update for a state fails to satisfy floors, or causality, we revert fluxes to $\widetilde{\mathcal{F}}^{\mathrm{LO}}$ (see \S\ref{sbsbsec:coll_src}). In Fig.\ref{fig:diffusion_test_mm_split} (lower) the effect of deactivating this form of flux-correction (FC) is investigated. At lower resolutions the trailing edge of the profile has points that acquire floor values without FC. In order to exclude \MeshBlock{} structure as a potential cause, we have verified that this behaviour is also present on domains without grid-partitioning. In order to preserve robustness, we therefore always run with M1(+N0) FC enabled.

Overall this test strongly suggests that our implicit treatment of the collisional source terms is sufficient to capture the advection of trapped radiation (see \cite{Radice:2021jtw}).

\subsection{Shadow casting}
\label{sbsec:shadow_test}
Strongly absorbing objects should cast a shadow when impinged upon by an illuminating, collimated beam. This is a property of the M1 scheme \cite{Fragile:2014bfa}, we now verify, based on a two-dimensional test \cite{sadowski2013semiimplicitschemetreating,Anninos:2020bpo,weih2020twomomentschemegeneralrelativistic,Radice:2021jtw,Izquierdo:2022eaz,Schianchi:2023uky,musolino2024practicalguidemoment}.

For this test, the \Mesh{} is set to be a thin slab, three-dimensional Cartesian grid. We take $x\in[-2,\,4]$, $y\in[-3,\,3]$ with $N_M=120$. The $z$ direction has \Mesh{} sampling consisting of four points, with extent set such that $\delta x=\delta y = \delta z$. Two levels of static refinement are added so as to resolve $(x,\,y)\in[-1,\,1]\times[-1,\,1]$ which contains an absorbing feature (see below). The time-step is chosen so as to satisfy a CFL condition of $0.25$. The background space-time is again assumed to be flat.

Initial data is selected such that $\widetilde{E}\big|_{t=0}=0=\widetilde{F}{}_i\big|_{t=0}$ on the interior of $\Omega$. To inject a collimated beam, with direction of propagation aligned with the $x$-axis, we set $\widetilde{E}\big|_{t\geq0,x=0,y\in[-1,\,1]}=1=\widetilde{F}{}_i\big|_{t\geq0,x=0,y\in[-1,\,1]}$. An outflow condition is imposed on the remaining faces of $\partial\Omega$. The absorbing object is taken to be a cylinder $\mathcal{C}$, centered at the origin, with radius $r=1$. Density is set to $\rho=1$. All opacities are taken as zero apart from within $\mathcal{C}$ where $\kappa{}_{\mathrm{a}}=1$ is set.

\begin{figure}[t]
  \centering
    \includegraphics[width=0.49\textwidth]{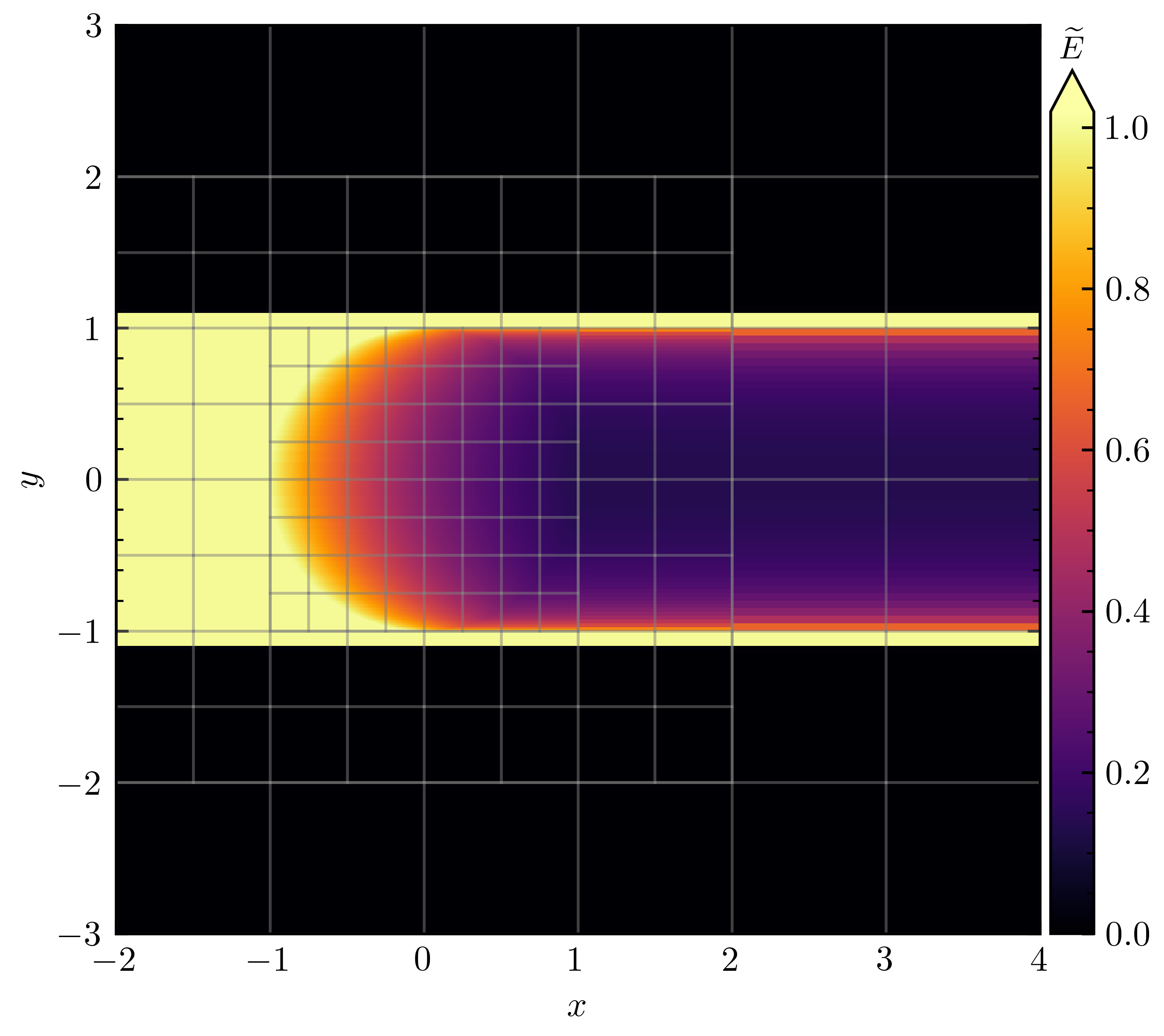}
    \caption{Shadow casting: Euler-frame energy density $\widetilde{E}$ at time  $t=10$. An incident beam is injected at the left boundary, and impinges upon an absorbing cylinder $\mathcal{C}$. Observe that the field remains regular as it propagates towards the two levels of refinement that contain $\mathcal{C}$, and exits towards the right.
    In gray we indicate \MeshBlock{} boundaries, where the interior of each has $N_B=(20,\,20,\,4)$ samples along each axis. Observe the static refinement structure of the grid towards the central feature.
    }
 \label{fig:shadow_test}
\end{figure}

In Fig.\ref{fig:shadow_test}, we depict the solution at time $t=10$, where steady-state has been attained. We do not observe significant lateral spreading or spurious formation of unsteady oscillation of the radiation field in the wake of the cylinder.

\subsection{Radiating sphere}
\label{sbsec:hom_sphere}

As an (extremely simplified) model of an isolated, radiating neutron star, with a geometry of typical of astrophysical applications, we consider a homogeneous radiating and absorbing sphere \cite{smit1997hyperbolicitycriticalpoints,Radice:2021jtw,Izquierdo:2022eaz,Schianchi:2023uky,musolino2024practicalguidemoment}.

The model consists of a static, spherically symmetric, homogeneous sphere, of constant energy density within radius $R$, embedded in a vacuum region, assumed to be radiating while in equilibrium. Define $r:=\sqrt{x^2+y^2+z^2}$, and fix $R=1$. We set $v{}^i\big|_{t=0}=0$, $\widetilde{E}\big|_{t=0,r\leq R} = 1$, with non-zero opacities for $r\leq R$ set according to $\mathcal{K}:=\kappa{}_{\mathrm{a}}=\eta$. The background space-time is assumed to be flat.

For the \Mesh{} we take $\Omega=[-4,\,4]^3$, with $N_M=64$ sample in each direction. Two levels of static refinement are added to resolve $[-1,\,1]^3$, where each \MeshBlock{} has $N_B=16$ samples in each direction. On $\partial\Omega$ outflow BC are set. The time-step is chosen so as to satisfy a CFL condition of $0.3$.

\begin{figure}[t]
  \centering
    \includegraphics[width=0.49\textwidth]{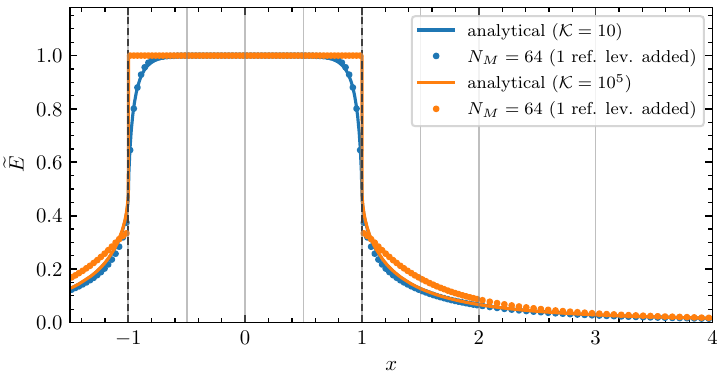}
    \caption{Radiating sphere: on-axis profile of Euler-frame energy density $\widetilde{E}$ at time $t=10$. Two choices for the opacities are selected through $\mathcal{K}$. The numerical M1 profiles compare favourably with the full analytical solution based on the Boltzmann equation. Vertical dashed black line indicates location of the spherical surface. In gray \MeshBlock{} boundaries are depicted. Observe the static refinement structure of the grid towards the central feature.}
 \label{fig:sphere_radabs_test}
\end{figure}

In the idealised setting assumed here an analytical solution to the Boltzmann equation can be constructed \cite{smit1997hyperbolicitycriticalpoints}. In Fig.\ref{fig:sphere_radabs_test} we compare the numerical result of our M1 implementation for two choices of $\mathcal{K}$ to the aforementioned. Excellent agreement is achieved for $\mathcal{K}=10$ whereas minor deviation is visible for $\mathcal{K}=10^5$. As noted in e.g.~\cite{Schianchi:2023uky}, convergence to the exact solution cannot be expected as M1, by construction, is an approximation scheme for the Boltzmann problem.

\subsection{Gravitational light bending}
\label{sbsec:light_bending}

As a test of our M1 implementation that involves strongly curved space-time, we consider propagation of a beam on a Schwarzschild background, where the metric is expressed in Cartesian Kerr-Schild coordinates \cite{mckinney2014threedimensionalgeneralrelativistic,Foucart:2015vpa,weih2020twomomentschemegeneralrelativistic,Radice:2021jtw,Schianchi:2023uky,musolino2024practicalguidemoment}. In particular, the gravitational sources $\widetilde{\mathcal{G}}{}_A$ appearing in Eq.\eqref{eq:m1n0_balance_law} are no longer trivial, and are handled numerically.

For this problem, the \Mesh{} is set to be a thin slab, where $(x,\,y)\in[0,\,5]^2$ with $N_M=100$. The $z$ direction has \Mesh{} sampling consisting of four points, with extent set such that $\delta x=\delta y = \delta z$. In addition, two levels of refinement are dynamically added in regions where $\widetilde{E}\geq 0.3$ develops. The time-step is chosen so as to satisfy a CFL of $0.2$.

The background fluid is taken to be completely transparent (zero opacities), and we position a unit-mass (code units) black-hole at the origin of the Cartesian grid. A unit-width $\widetilde{E}=1$ beam, centered at $x=0$, $y=3.5$, is continually injected. The incident direction of propagation is taken as the $x$-direction. To this end, we set $\widetilde{F}{}^i$ such that $\alpha \widetilde{F}{}^i - \beta{}^i \widetilde{E}$ is along the $x$-axis, and $\Vert\widetilde{F}\Vert_\gamma = 0.99\widetilde{E}^2$. On the remaining faces of $\partial\Omega$ outflow BC are set.

\begin{figure}[t]
  \centering
    \includegraphics[width=0.49\textwidth]{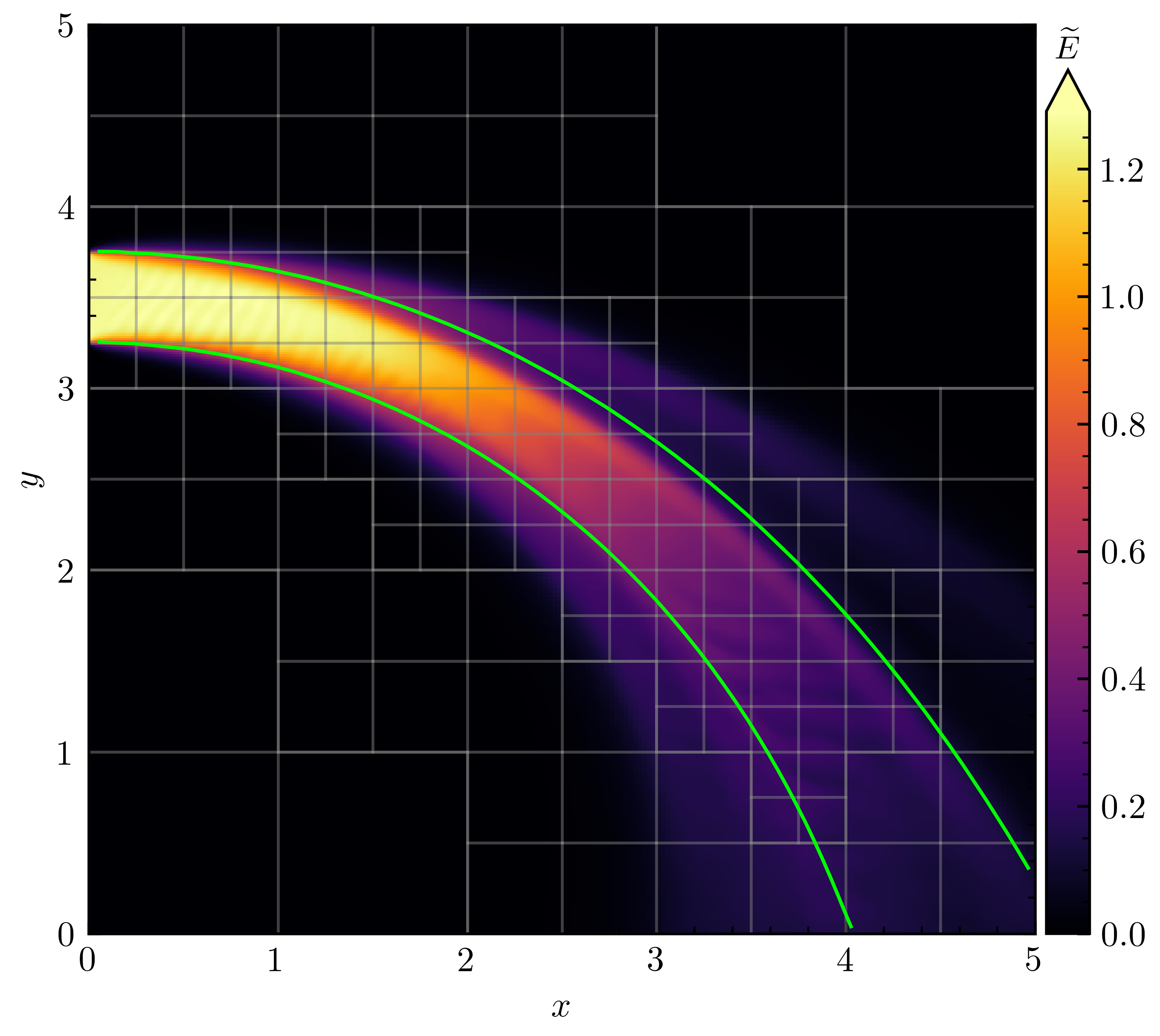}
    \caption{Gravitational light-bending: Euler-frame energy density $\widetilde{E}$ at time $t=20$. The radius of the black hole is set as $R_{\mathrm{BH}}=2$. Geodesics are shown in lime-green. The trajectory of the beam indicates that behaviour under non-zero curvature is handled correctly.
    Gray lines indicate {\MeshBlock{}} boundaries. Observe that AMR structure has dynamically developed to capture the high intensity $(\widetilde{E}\geq 0.3)$ region of the beam.  }
 \label{fig:Kerr_beam_test}
\end{figure}

In Fig.\ref{fig:Kerr_beam_test} we show the energy density at time $t=20$, after steady state has been attained. The trajectory of $\widetilde{E}$ is well-captured by the null geodesics of the geometric background. While some lateral diffusion is present, this is a numerical artifact, also appearing in other M1 implementations \cite{mckinney2014threedimensionalgeneralrelativistic,Foucart:2015vpa,weih2020twomomentschemegeneralrelativistic,Radice:2021jtw,Schianchi:2023uky,musolino2024practicalguidemoment}.

\subsection{Opacity test: cross-code comparison}
\label{sbsec:eql_handling}

The opacity library we have ported for this work is \texttt{weakrates} of \THC{} \cite{Radice:2018ghv,Radice:2021jtw} (see \S\ref{sbsec:opacities}). The test problems investigated here, thus far, have not involved this functionality but instead directly specified opacities. To close this gap, with a final, simple, sanity-check, we proceed as follows. An initial, hydrodynamical state $\mathcal{H}$, characterised by $(\rho,\,T,\,Y{}_e)$, together with EOS is selected. This choice is used to seed a translation-invariant, periodic $\Omega$. The time-evolution of the state $\mathcal{H}(t)$, as evolved by our implementation in \GRAthena{}, is compared with that of \THC{}.

For this test we choose the (tabulated) DD2 EOS \cite{Typel:2009sy,Hempel:2009mc}. The fluid velocity is taken to be $v_x=0.7$. The \Mesh{} is comprised of uniform, unit grid spacing, and $N_M=4$ samples in each direction. The time-step is selected to satisfy a CFL of $0.25$. The background space-time is taken to be flat.

\begin{figure}[t]
  \centering
    \includegraphics[width=0.49\textwidth]{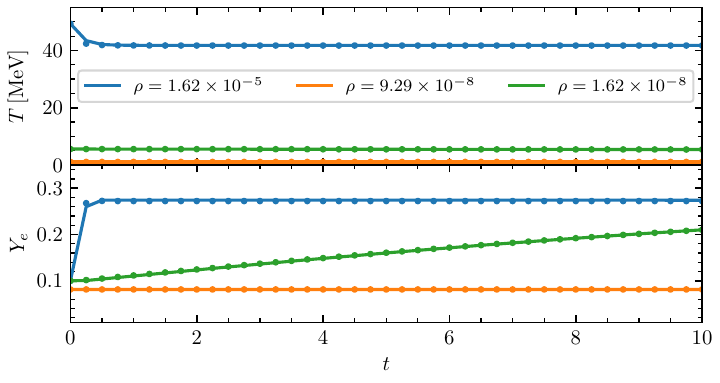}
    \caption{Opacity test: evolution of quantities at a central cell with \GRAthena{} as depicted in solid lines, and seen as consistent with that of \THC{} as shown in filled circles. In the upper panel temperature is shown, whereas in the lower panel electron fraction. Behaviour is inspected for densities indicated in the inset where (blue) models near-trapped (optically thick) conditions; (green) free-streaming regime; (orange) low-temperature conditions that occur in (e.g.)~the vicinity of a remnant.}
 \label{fig:eql_test}
\end{figure}

In Fig.\ref{fig:eql_test} the result of this test is shown for three choices of initial $\mathcal{H}$ that probe: optically thick, free-streaming, and low-temperature conditions. In each case we find that excellent consistency is maintained across codes. We have verified that this agreement persists to (at least) a final time of $t=100$. This strongly suggests equilibrium conditions (when relevant) are correctly preserved.

We emphasize that while simple this test is particularly stringent as it is focused on the interplay between $\widetilde{N}{}_{(\nu)}$ evolution, fluid sector coupling, opacities supplied by \texttt{weakrates}, and equilibrium condition handling (see e.g.~\cite{Perego:2019adq}), which \GRAthena{} passes.

\section{Applications}
\label{sec:applications}

We now activate the full GR(M)HD-M1+N0 capabilities of \GRAthena{} and study physical scenarios involving rotating collapse of isolated neutron stars (NS) \S\ref{sbsec:rot_collapse}, followed by binary neutron star (BNS) mergers \S\ref{ssec:bns_calib}.

The grid over the computational domain $\Omega$ is assembled, and refined, based on the strategy of \cite{Daszuta:2024ucu}: fix a tracker $\mathbf{c}(t)$ to an NS constituent that follows the minimum of the lapse; impose there a desired level of refinement, over a centered ball $\mathcal{B}_r(t;\,\mathbf{c})$ so as to achieve a target resolution.

For both problems $\mathcal{B}_r(t;\,\mathbf{c})$ is selected to have radius $22.1\,[\mathrm{km}]$. In the case of rotating collapse, we set $x_D = 756.0\,[\mathrm{km}]$. In the case of a BNS, we work with two grids, one with reduced spatial extent for calibrating our simulations, where $x_D=567\,[\mathrm{km}]$, and a production grid, where $x_D=2268.1\,[\mathrm{km}]$. Grids featuring distinct resolution are constructed by selecting an $N_M\in\{96,\,144\}$ and adding refinement over $\mathcal{B}_r(t;\,\mathbf{c})$ until a target resolution is reached. We emphasize that adding refinement as mentioned induces a global change in the \Mesh{} due to the imposition of (at most) $2:1$ refinement-ratio between nearest-neighbor \MeshBlock{} objects. For the runs here, we consider a collection of three finest-level resolutions  $\mathcal{T}_\rho:= (\delta x_c,\,\delta x_m,\,\delta x_f)=(369.2,\, 246.1,\,184.6)\,[\mathrm{m}]$. Unless otherwise stated, the CFL is taken to be $0.25$.

\subsection{Diagnostic quantities}
\label{sbsec:diagnostics}

As a preliminary, recall that the flux integral $\Phi$, over a 2-surface $\Sigma$, of the $\sqrt{\gamma}$-densitized flux $\widetilde{\mathcal{F}}$ can be defined through the expression $\Phi:=\int_\Sigma \widetilde{\mathcal{F}}{}_i\,\mathrm{d}\Sigma^i$. In particular, for a spherical surface $\mathbb{S}^2_r$, of radius $r$, we set $\Sigma=\mathbb{S}^2_r$, with $\mathrm{d}\Sigma^i:=\alpha r^2 \hat{r}^i \,\mathrm{d}\Omega$, where $\mathrm{d}\Omega$ is the 2-sphere area element. In the case of neutrino luminosity $\Phi_{L_\nu}:=L{}_{\nu}$, we set the flux as $\widetilde{\mathcal{F}}{}^i_1$ as in Eq.\eqref{eq:m1flxvec}. For the (per-species) average energy we evaluate $\langle \varepsilon_\nu\rangle:= \Phi_{L_\nu}/\Phi_{\mathcal{N}_\nu}$, where $\Phi_{\mathcal{N}_\nu}$ is an integral involving neutrino number flux $\widetilde
{\mathcal{F}}{}^i_0$.

Detailed properties of the ejecta, such as its mass, composition, geometry, and entropy are important to analyse for understanding the physical properties of the remnant. We identify matter as unbound according to two common criteria: geodesic as characterized by $u_t<-1$ \cite{Hotokezaka:2012ze}, and Bernoulli in the form (see \cite{Fujibayashi:2020dvr,Foucart:2021ikp}) $h u_t < -h_{\mathrm{min}}$ where $h$ is the specific-enthalpy and $h_{\mathrm{min}}$ is the minimum allowed value in a given tabulated EOS. Define the flagged conserved density $\widetilde{D}_g$ such that $\widetilde{D}_g=\widetilde{D}$ when the geodesic criterion is satisfied and $0$ otherwise, and similarly $\widetilde{D}_b$ in the Bernoulli case. By assembling the fluid flux $\widetilde{\mathcal{F}}{}^i=\widetilde{D}(\alpha v{}^i-\beta{}^i)$, and imposing strict radial outflow ($\widetilde{\mathcal{F}}_r>0$), we compute the ejected mass based on either criterion. Weighted-averages are computed according to $\langle X \rangle:= \int_\Sigma X\,\widetilde{\mathcal{F}}_i\,\mathrm{d}\Sigma{}^i\,/\Phi$.

During the course of a simulation, dynamical fields are interpolated onto origin-centered, concentric spherical shells, arranged in a sequence of increasing radius, and sampled in equi-angular fashion. The flux integrals under discussion here, are evaluated via standard numerical quadrature in a post-processing step. Unless otherwise stated, we present results utilizing an extraction radius of $r\simeq 295\,[\mathrm{km}]$.

\subsection{Rotating collapse}
\label{sbsec:rot_collapse}

To test the stability of the code during the formation of a black hole, we simulate the collapse of an initially cold, uniformly rotating, magnetized neutron star using the SFHo EOS \citep{Steiner:2012rk}.
We calculate the initial data with the \texttt{RNSC} code \citep{Stergioulas:1994ea} with a central density of ${1.6314} \times 10^{15}\,[ \mathrm{g}\, \mathrm{cm}^{-3}]$ and a polar-to-equatorial coordinate axis ratio of $r_p/r_e = 0.55$. 
This configuration leads to star with rotational frequency $10.615\,[\mathrm{kHz}]$,
baryon mass $M_b = 2.86(38)\,[M_\odot]$, gravitational mass $M = 2.44(54)\,[M_\odot]$, and dimensionless spin parameter $\chi=0.677$, which is close to the maximum-mass, maximum-rotation configuration.
A poloidal magnetic field with maximum strength $5\times10^{15}\,[\mathrm{G}]$ is superposed in a way similar to what is described in Sec.~\ref{ssec:bns_calib} below.

\begin{figure}[t]
  \centering
  \includegraphics[width=0.49\textwidth]{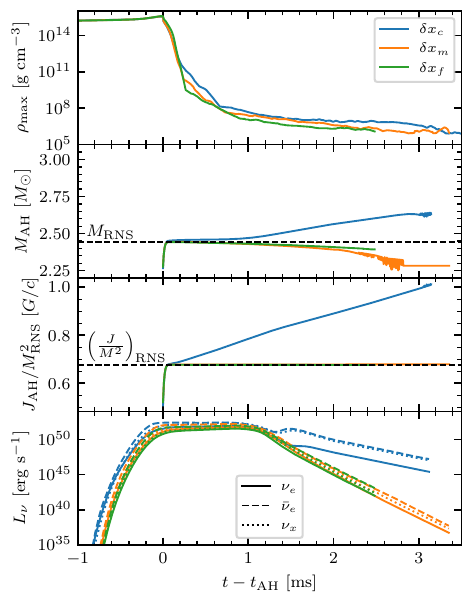}
  \caption{Maximum density, horizon mass, horizon spin, and neutrino luminosities for the rotating neutron star collapse at 3 resolutions.
      The mass and angular momentum of the initial data are indicated by black dashed lines.}
  \label{fig:RNS_scalars}
\end{figure}

We perform GRMHD+M1+N0 simulations for the resolutions $\delta x_c,\,\delta x_m,\,\delta x_f\in\mathcal{T}_{\rho}$ until $2$--$3\,[\mathrm{ms}]$ after horizon formation. The star immediately collapses due to the perturbation introduced by truncation errors and initial data interpolation errors. Magnetic pressure cannot prevent the collapse and the feedback of the magnetic field dynamics on the fluid is practically negligible. A light disc forms and entirely accretes onto the black hole by the end of the simulation. 
Figure~\ref{fig:RNS_scalars} shows the evolution of the maximum density, horizon mass, horizon spin, and neutrino luminosities (bottom panel) for the 3 resolutions.
An apparent horizon is found at $t_{\text{AH}} \simeq 1.246, 1.187, 1.170\,[\mathrm{ms}]$ for $\delta x_c$, $\delta x_m$, and $\delta x_f$, respectively.
As soon as the horizon is formed, the star is accreted onto the black hole within ${\sim} 0.2\,[\mathrm{ms}]$.
The magnetic, neutrino, and matter fields are excised by the scheme described in \S\ref{sbsec:ahf_excision_tap} within 55\% of the horizon radius ($f_r=0.55$).
As a result, the maximum density rapidly drops after an apparent horizon is found.

After the horizon forms, the horizon mass and spin quickly saturate around $M\simeq2.45 \,[M_\odot]$ and $\frac{c J}{G M^2}\simeq0.677$
These values match the corresponding values of the initial data since the energy and angular momentum from the gravitational waves are within truncation errors, Cf.~\cite{Reisswig:2012nc,Dietrich:2014wja}.
In the coarse resolution simulation ($\delta x_c$), the found apparent horizon mass and spin continue to grow unphysically, and the neutrino luminosities are not decreasing as expected, likely due to insufficient resolution. At mid and fine resolution, the horizon mass is approximately constant except for a slow drift towards lower mass by the end of the simulation time.
About ${\sim} 2.5\,[\mathrm{ms}]$ after the horizon formation, the horizon-finder solution becomes slightly unstable at mid-resolution $\delta x_m$. We tested this can be significantly improved by adding a refinement level at horizon formation, see Sec.~\ref{sssec:sfho_excision}. In the fine-resolution simulation the horizon stays stable for the duration of the simulation.
At both $\delta x_m$ and $\delta x_c$, the neutrino luminosity starts decreasing ${\sim} 1\,[\mathrm{ms}]$ after the horizon is formed (roughly the time-of-flight delay to the detection sphere at $r\approx c \times 1$\,ms) as neutrinos and baryonic matter are swallowed by the newly formed black hole.

\begin{figure*}[t]
  \centering
  \includegraphics[width=\textwidth]{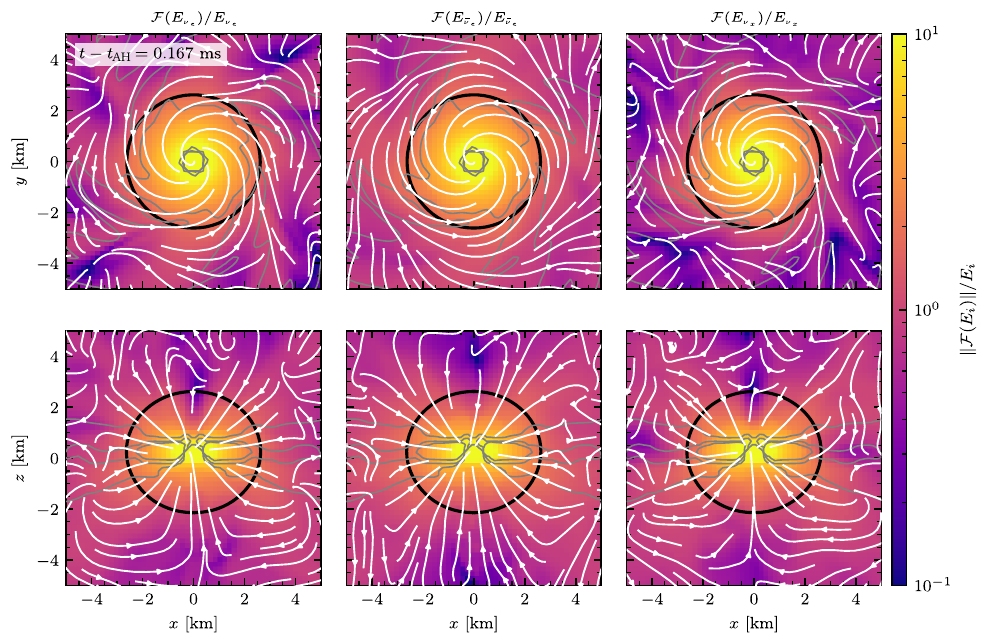}
	\caption{Neutrino energy fluxes normalized to the neutrino energy density 0.167\,ms after the formation of the apparent horizon.
    Top and bottom panels show the $xy$ and $xz$ plane, respectively.
    Left, middle, and right panels show $\nu_e, \bar{\nu}_e$, and $\nu_x$, respectively.
    Colors indicate the absolute values of the flux and arrows the direction.
    The black circle shows the position of the apparent horizon and the gray contours show the density isosurfaces at $\rho \in \{10^{8},10^{9}, 10^{10}\}\,\text{g\,cm}^{-3}$.}
  \label{fig:RNS_F_nu}
\end{figure*}

Figure \ref{fig:RNS_F_nu} shows the neutrino-energy-density fluxes normalized by the neutrino energies for all three evolved neutrino species in the left, middle, and right panels, respectively. The black circle shows the position of the apparent horizon; the $(x,y)$-plane refer to the plane perpedicular to the rotation axis.
The neutrino flow (illustrated by white streamlines) exclusively points into the horizon surface (shown by black lines) meaning that no neutrinos escape the horizon.
Therefore, our excision scheme allows robust evolutions of the radiation during and beyond black hole formation.

\subsection{Binary neutron stars}\label{ssec:bns_calib}

We now consider modeling equal-mass binary neutron stars (BNS). The approach adopted here is two-step. First we calibrate our simulation based on initial data (ID) constructed at tighter separation (resulting in fewer orbits), on a grid of reduced overall spatial extent $x_D=567.0\,[\mathrm{km}]$. This has the advantage of allowing us to more rapidly iterate upon investigating the particulars of solver choice, at reduced overall computational cost. The reduced separation and grid simulations are complemented by runs involving larger production grids, and ID at greater separation in \S\ref{ssec:bns_prod}. Importantly, we do not impose any underlying symmetry for the runs presented.

To model the BNS, ID is prepared utilizing the \texttt{Lorene} library \cite{Gourgoulhon:2000nn}. We consider irrotational (non-spinning) neutron star (NS) constituents in a quasi-circular orbit, under the assumption that matter is in beta-equilibrium with a constant initial temperature of $T=0.1\,[\mathrm{MeV}]$. 
In this section, we employ the DD2 EOS \cite{Typel:2009sy,Hempel:2009mc}, and select ID such that a long-lived remnant ensues. The constructed ID features NS component masses of $M_1=M_2=1.296\,[\Msun]$, and baryonic masses $M_{1b}=M_{2b}=1.350\,[\Msun]$. The system has an ADM mass of $M{}_{\mathrm{ADM}}\simeq 2.672\,[\Msun]$. The initial ADM angular momentum of the BNS system is $J{}_{\mathrm{ADM}}\simeq 6.957\,[\Msun^2]$.

For runs that additionally include (poloidal) magnetic fields, we follow the approach of our recent work \cite{Gutierrez:2025gkx,Cook:2025frw}. A purely toroidal vector potential is prescribed through $\mathbf{A}=A{}_b\max(p-p{}_c,\,0)(-y,\,x,\,0)$, with $p{}_c=0.04p{}_{\mathrm{max}}$, where for each star $p{}_{\mathrm{max}}$ is the maximum pressure, and $(x,\,y)$ is measured relative to the center of an NS constituent. Unless otherwise stated, the free parameter $A{}_b$ is chosen such that conserved ($\sqrt{\gamma}$-densitized) magnetic field satisfies $\widetilde{\mathcal{B}}{}_{\mathrm{max}}\big|_{t=0} =5\times 10^{15}\,[\mathrm{G}]$, where local maxima are located at the center of each NS. 
Typically \cite{Foucart:2015gaa,Radice:2021jtw,Zappa:2022rpd,Schianchi:2023uky,Espino:2023mda,Cheong:2024buu} for simulations involving M1(+N0), the (magneto)-hydrodynamical sector is treated with an HLLE solver at minimum. We contrast qualitative features involving this choice, against that of LLF in what follows.

\subsubsection{Overview of dynamics and (M)HD solver impact}\label{sssec:overview_reduced}

For the $39\,[\mathrm{km}]$ separation DD2 binaries we find that they undergo ${\sim} 3$ orbits, prior to merger\footnote{Time of merger is taken as the time of peak amplitude in the $\ell=m=2$ mode of the gravitational wave strain.}. The time of merger occurs at $t_{\mathrm{mrg}}\simeq 6.4\,[\mathrm{ms}]$ irrespective of (magneto)-hydrodynamical solver choice, and whether or not magnetic fields are present.

\begin{figure*}[t]
  \centering
    \includegraphics[width=\textwidth]{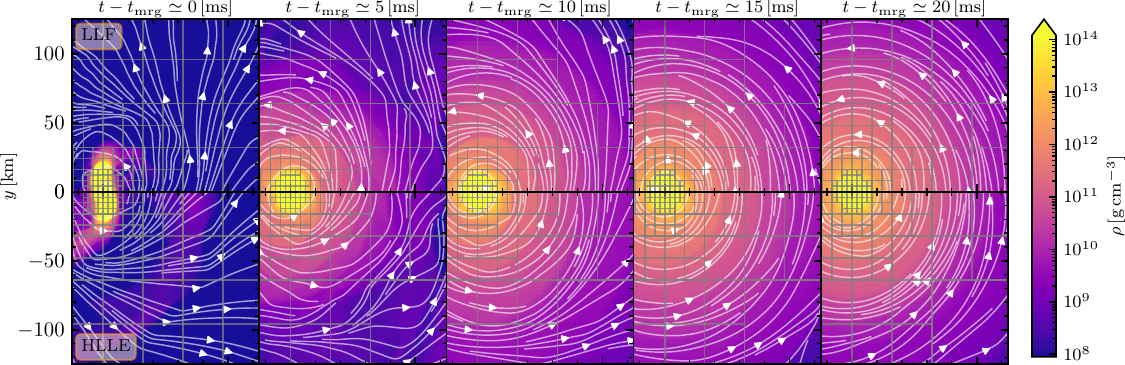}
    \includegraphics[width=\textwidth]{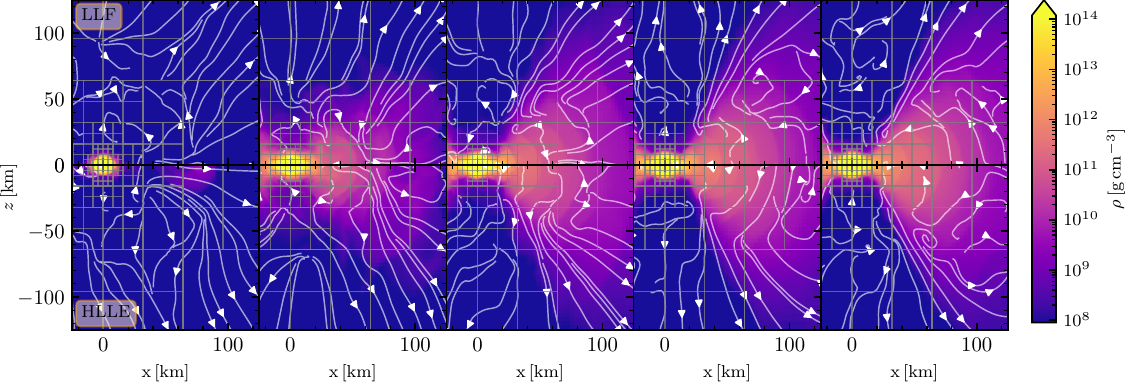}
    \caption{Representative snap-shots of the density $\rho$ of the DD2 binary as it evolves in time. These runs involve magnetic fields and M1+N0. In the upper sub-plot $x$-$y$ equatorial slices are shown at time of merger, and spaced by $5\,[\mathrm{ms}]$ towards the right thereafter, to a final time of $20\,[\mathrm{ms}]$. In the lower sub-plot $x$-$z$ meridional slices are shown. Each sub-plot has an upper and lower panel that contrasts the evolution utilizing LLF and HLLE solvers for the hydrodynamical sector respectively. 
    Stream-lines (white) depict integral curves of the fluid velocity $v^i$ components in the salient plane under consideration. Gray lines indicate {\MeshBlock{}} boundaries, the structure of which has developed according to AMR criterion. See text for further discussion.}
 \label{fig:slice_rho_vel_mhd}
\end{figure*}

Representative snapshots, that commence at merger with $5\,[\mathrm{ms}]$ spacing thereafter, of the rest-mass density $\rho$ are displayed in Fig.~\ref{fig:slice_rho_vel_mhd}, contrasting LLF and HLLE solvers in both equatorial ($x$-$y$) and meridional ($x$-$z$) planes, for runs involving MHD. Flow stream-lines illustrate the fluid velocity field $v{}^i$.
At the time of merger we see that the HLLE solver supports a more intricate density structure which are absent in the LLF case.
This may be attributable to the lower diffusion of the HLLE solver permitting the tidal ejecta to separate from the remnant with less numerical mixing. 
Overall, however, qualitatively similar structure develops in the later post-merger phase for both solvers. The AMR block structure, demarcated in gray lines, adapts similarly in both cases. We note that these observations carry over for runs in the absence of magnetic fields also.

\begin{figure*}[t]
  \centering
    \includegraphics[width=\textwidth]{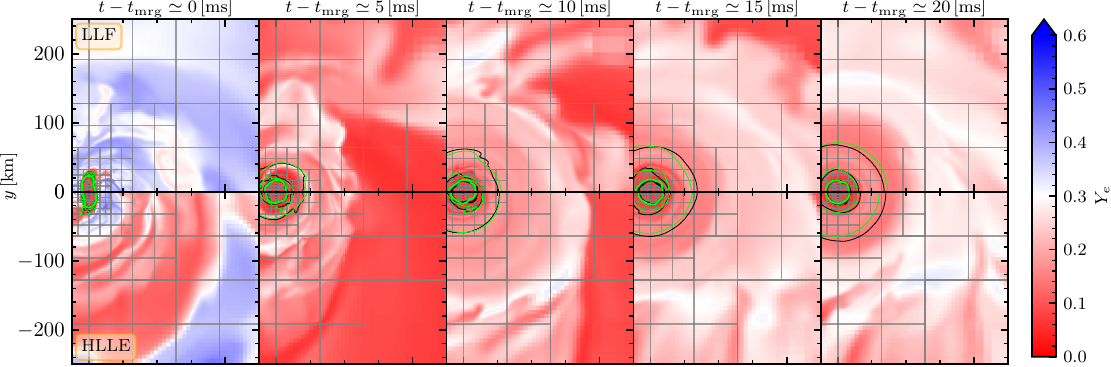}
    \includegraphics[width=\textwidth]{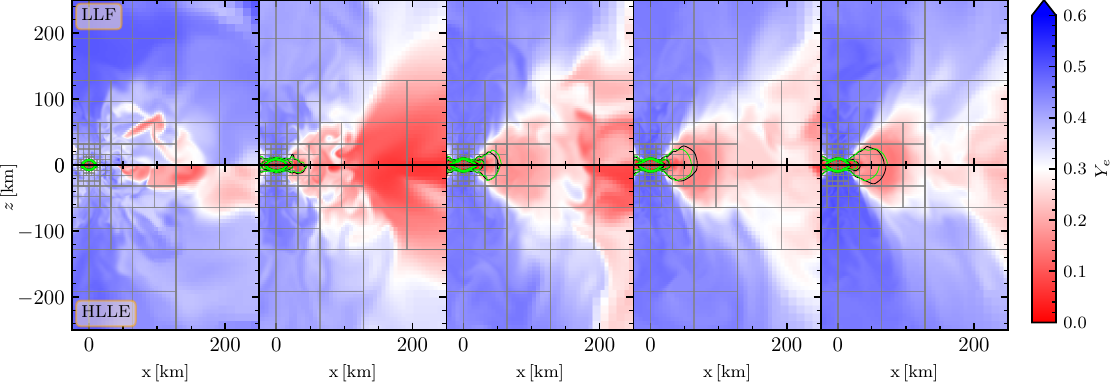}
    \caption{Representative snap-shots of the electron fraction $Y_{\mathrm{e}}$ of the DD2 binary as it evolves in time. These runs involve magnetic fields and M1+N0. In the upper sub-plot $x$-$y$ equatorial slices are shown at time of merger, and spaced by $5\,[\mathrm{ms}]$ towards the right thereafter, to a final time of $20\,[\mathrm{ms}]$. In the lower sub-plot $x$-$z$ meridional slices are shown. Each sub-plot has an upper and lower panel that contrasts the evolution utilizing LLF and HLLE solvers for the hydrodynamical sector respectively.
    Isodensity contours where nested black lines indicate $\rho\in\{10^{11},\,10^{12},\,10^{13}\}\,[\mathrm{g}\,\mathrm{cm}^{-3}]$ for runs with magnetic field; contours in light-green depict equivalent runs in the absence of magnetic fields. Gray lines indicate {\MeshBlock{}} boundaries the structure of which has developed according to AMR criterion. See text for further discussion.}
 \label{fig:slice_Y_e_mhd}
\end{figure*}

The corresponding evolution of the electron fraction $Y_{\mathrm{e}}$ is shown in Fig.~\ref{fig:slice_Y_e_mhd}. It is clear that shortly post-merger the simulation with HLLE supports generation of a more neutron rich outflow. In the post-merger remnant, neutrino emission and absorption drive $Y_e$ evolution: the dense core remains neutron-rich ($Y_{\mathrm{e}}\lesssim 0.1$), while shock-heated material at lower densities ($\rho\lesssim 10^{13}\,[\mathrm{g}\,\mathrm{cm}^{-3}]$) develops higher $Y_{\mathrm{e}}$. Isodensity contours from MHD runs (black) are overlaid with those from equivalent HD runs (light-green). Minor differences are apparent in the overall extent of the remnant disc. These drive towards a late-time state (not-depicted) where simulations involving the HLLE solver lead to a slightly more compact disc, when contrasted against LLF (see also \S\ref{sssec:lep_neutr}).

\begin{figure}[t]
  \centering
    \includegraphics[width=0.49\textwidth]{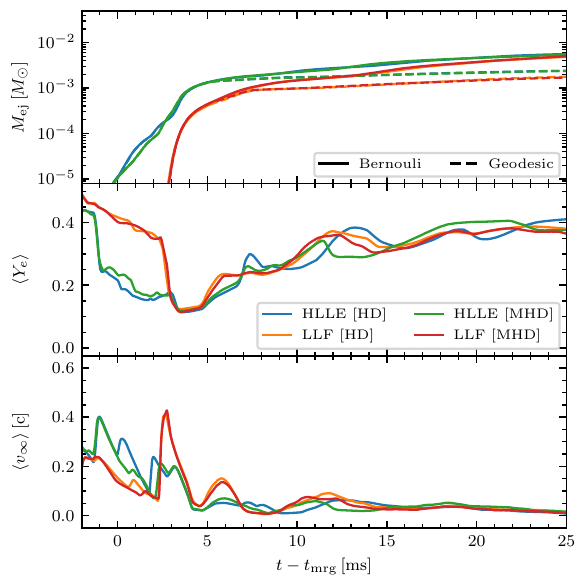}
    \caption{(Upper panel) Cumulative ejected mass $M_{\mathrm{ej}}$ according to Bernoulli and geodesic criteria in solid and dashed lines respectively.
    (Middle panel) Average electron fraction $\langle Y_{\mathrm{e}}\rangle$ passing through extraction sphere.
    (Lower panel) Average asymptotic velocity $\langle v_\infty \rangle$ passing through extraction sphere.
    (Common) Resolution of $\delta x_c$ has been fixed. (M)HD together with matter solver employed, are as indicated by the legend of the middle panel. Extraction sphere of radius $R\simeq 295\,[\mathrm{km}]$ is utilized. See text for discussion.
    }
 \label{fig:DD2_reduced_ej_m_y_v}
\end{figure}

Cumulative ejecta mass $M_{\mathrm{ej}}$, average electron fraction $\langle Y_{\mathrm{e}}\rangle$, and average asymptotic velocity $\langle v_\infty\rangle$ are compared across solver and (M)HD choice at fixed resolution in Fig.~\ref{fig:DD2_reduced_ej_m_y_v}. Here, the asymptotic velocity is defined as $v_\infty:=\sqrt{1-1/(h\,u_t)^2}$, where $h$ is the specific enthalpy and $u_t$ is the time component of the four-velocity. Systematic differences between LLF and HLLE are clearly apparent. In particular, the cumulative ejecta mass $M_{\mathrm{ej}}$ differs between solvers, with an earlier onset present in HLLE simulations. We observe that both HLLE and LLF appear to be approaching a common trend at $t-t_{\mathrm{mrg}}\sim 25\,[\mathrm{ms}]$. Regarding electron fraction, the HLLE runs exhibit an earlier ejection of neutron-rich material (Cf.~Fig.\ref{fig:slice_Y_e_mhd}), resulting in $\langle Y_{\mathrm{e}}\rangle\sim 0.2$ on the window $t-t_{\mathrm{mrg}}\in[-1,\,4]\,[\mathrm{ms}]$, substantially lower than the $\langle Y_{\mathrm{e}}\rangle\sim 0.39$ found for LLF. Subsequent to $t-t_{\mathrm{mrg}}\simeq 4\,[\mathrm{ms}]$ both solvers lead to a common trend, of gradual increase in $\langle Y_{\mathrm{e}}\rangle$. Finally, we note that $\langle v_\infty \rangle$ peaks for the HLLE cases $\sim 4\,[\mathrm{ms}]$ prior to the LLF runs at a value of $\langle v_\infty\rangle \sim 0.4\,[c]$.

Note that an element of caution must be employed in the discussion of ejecta for these runs on account of the comparatively small spatial extent ($x_D=567\,[\mathrm{km}]$). Consequently, small extraction sphere radii must be selected, and ejecta properties may be more prone to systematic boundary effects.
Relatedly, we emphasize the calibration runs of this section are at a relatively low resolution. Based on~\cite{Zappa:2022rpd}, a representative example matching the resolutions employed in our work can be extracted: the value of $M{}_{\mathrm{ej}}$ according to Bernoulli criterion, and following from an equal-mass BNS merger simulation,
and finest level resolution run with $\delta x_m$, is a factor of ${\sim} 2$ smaller at $30\,[\mathrm{ms}]$ post-merger when compared to a run based on the coarser $\delta x_c$, with the general trend indicating that higher resolution tends to depress $M{}_{\mathrm{ej}}$.


\subsubsection{Lepton fraction and neutrino surfaces}\label{sssec:lep_neutr}

\begin{figure}[t]
  \centering
    \includegraphics[width=0.49\textwidth]{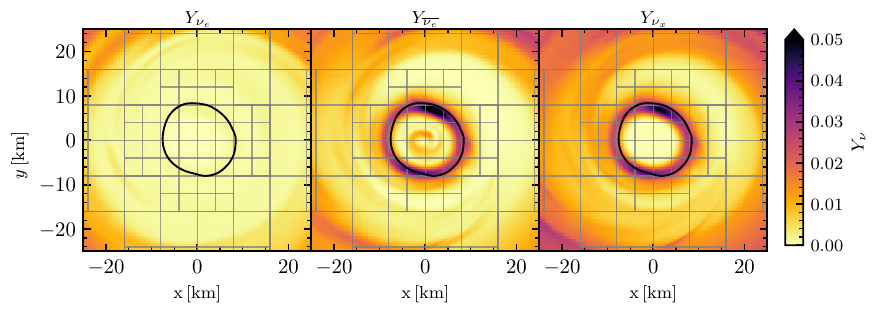}
    \caption{Lepton fractions $Y_\nu$ at $t-t_{\mathrm{mrg}}\simeq 20\,[\mathrm{ms}]$ for HLLE run with MHD. (Left-most panel) shows $Y_{\nu_\mathrm{e}}$, (middle panel) shows $Y_{\overline{\nu_\mathrm{e}}}$, and (right-most panel) shows $Y_{\nu_\mathrm{x}}$. Within the remanant core, we observe a hierarchy of $Y_{\nu_e}<Y_{\nu_x}<Y_{\nu_{\overline{\nu_e}}}$. This property is robust if instead HD is inspected, or the hydrodynamical treatment is changed to LLF. Isodensity contour (coloured black) indicates $\rho=10^{14}\,[\mathrm{g}\,\mathrm{cm}^{-3}]$.}
 \label{fig:slice_Y_nu_mhd}
\end{figure}

The impact of local thermodynamical conditions of the remnant on neutrino species fraction is to induce a hierarchy $Y_{\nu_e}<Y_{\nu_x}<Y_{\nu_{\overline{\nu_e}}}$ (\cite{Foucart:2015gaa,Perego:2019adq,Zappa:2022rpd}) within the interior, high-temperature, high density ($\rho\geq 10^{14}\,[\mathrm{g}\,\mathrm{cm}^{-3}]$) region. We show this directly for an HLLE run with MHD at $20\,[\mathrm{ms}]$ post-merger in Fig.\ref{fig:slice_Y_nu_mhd}. We have verified that this remains robust in the absence of magnetic fields, and under change of MHD solver. The abundance of each species, can be understood based on thermodynamical conditions in the remnant NS core ~\cite{Perego:2019adq,Zappa:2022rpd}.

\begin{figure}[t]
  \centering
    \includegraphics[width=0.49\textwidth]{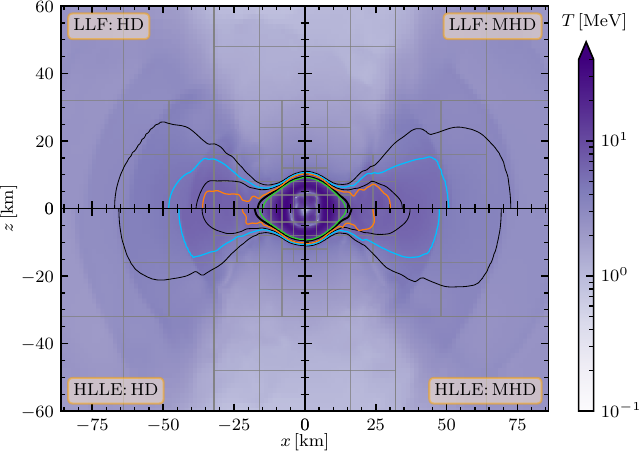}
    \caption{Neutrino surfaces at $t-t_{\mathrm{mrg}}\simeq 65\,[\mathrm{ms}]$ in $x$-$z$ meridional plane computed fixing $\tau=2/3$ based on the corrected opacities. For each quadrant a distinct choice of (M)HD and hydrodynamical solver is made, as indicated by the inset. In all cases we observe a robust hierarchy in minimal distance $R$ to the neutrino surface, of a given species. Specifically $R_{\nu_e}>R_{\overline{\nu_e}}>R_{\nu_x}$, are depicted in blue, orange, green respectively. Differences in $R$ between between schemes and (M)HD are apparent. Isodensity contours in solid black depict $\rho\in\{10^{11},\,10^{12},\,10^{13}\}\,[\mathrm{g}\,\mathrm{cm}^{-3}]$.}
 \label{fig:slice_xz_neutrinosph}
\end{figure}

Neutrino surfaces of the remnant at $t-t_{\mathrm{mrg}}\simeq 65\,[\mathrm{ms}]$ are approximated\footnote{In general this is a non-trivial problem, see e.g.~\cite{Endrizzi:2019trv}.} based on the location of $\tau=\sqrt{\kappa{}_{\mathrm{a}}(\kappa{}_{\mathrm{a}}+\kappa{}_{\mathrm{s}})}=2/3$, and compared in the presence (absence) of magnetic fields, and for switching between LLF and HLLE solver in the $x$-$z$ meridional plane in Fig.\ref{fig:slice_xz_neutrinosph}. These are super-posed over the remnant temperature profile. If we denote the minimal distance to the neutrino surface by $R_\nu$, then we observe the hierarchy $R_{\nu_e}>R_{\overline{\nu_e}}>R_{\nu_x}$. This appears robust irrespective of (M)HD and hydrodynamical solver choice. We remark that the hierarchy is consistent with that observed in the absence of magnetic fields \cite{Schianchi:2023uky}, for a different EOS, though the relative magnitudes of $R_\nu$ differ. We do however observe systematic differences in spatial extent. Specifically, LLF (when compared with HLLE) appears to give rise to larger $R_\nu$ for all species. For HLLE, we observe that the inclusion of magnetic fields tends to marginally increase $R_\nu$, whereas the opposite occurs for LLF. Changes in the spatial extent can be understood by noting that neutrino surfaces are strongly determined by the density profile inside the remnant disc \cite{Endrizzi:2019trv}, which is compatible with the differences in remnant disc compactness between solvers noted in \S\ref{sssec:overview_reduced}.

Based on the post-processing analysis of \cite{Chiesa:2024lnu}, we note that incorporating inelastic (neutrino-electron/positron) scattering could have a substantial effect on the absolute values of $R_\nu$.

\subsubsection{Magnetic field diagnostics}\label{sssec:mag_reduced}

We now inspect properties of the magnetic fields for the reduced grid (extent) runs. To this end, we perform additional runs featuring finest level resolution of $\delta x_m$. Furthermore, to test robustness, we supplement the initial poloidal magnetic field data involving $\widetilde{\mathcal{B}}_{\mathrm{max}}\big|_{t=0}=5\times 10^{15}\,[\mathrm{G}]$ with additional runs that have amplitude increased by a factor of ten, which we term the ``strong-mag'' variant.

At the relatively coarse resolutions $\delta x_c$ and $\delta x_m$, we have tested on the reduced grid that the Kelvin--Helmholtz and magneto-rotational instability (MRI) modes, which drive rapid amplification of the magnetic field, are under-resolved \cite{Gutierrez:2025gkx}, which entails that the maximum field strength that can develop dynamically at early-to-intermediate post-merger times is limited relative to what higher resolution studies obtain~\cite{Kiuchi:2017zzg,Aguilera-Miret:2020dhz,Palenzuela:2021gdo,supp:Gutierrez:2026ngt}. Aspects of the consistency of the field evolution can however be assessed.

\begin{figure}[t]
  \centering
    \includegraphics[width=0.49\textwidth]{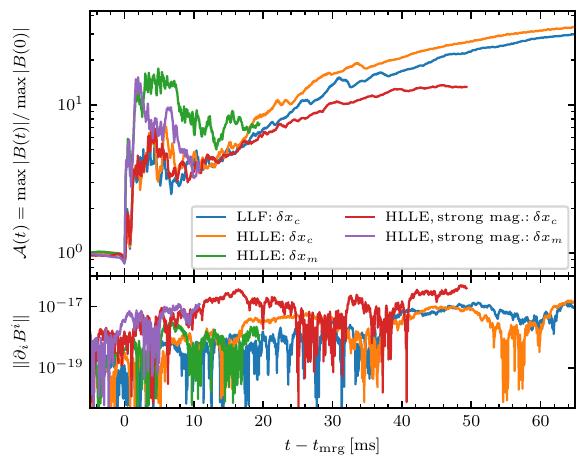}
    \caption{(Upper panel): amplification of maximum magnetic field strength relative to the maximum initial value for distinct choices of solver, finest resolution, and field strength. (Lower panel): Preservation of divergence-free constraint. See text for discussion.}
 \label{fig:BfieldAmpDiv_reduced}
\end{figure}

In Fig.\ref{fig:BfieldAmpDiv_reduced}, we display the evolution of the amplification factor $\mathcal{A}(t)$ defined as the maximum magnetic field strength relative to its maximum initial value (upper panel) and the preservation of the divergence-free constraint (lower panel) for the reduced-grid MHD runs. We observe a rapid increase in $\mathcal{A}(t)$ at the onset of merger. Considering first the standard field strength runs at a fixed finest level resolution of $\delta x_c$, we find that, for LLF, $\mathcal{A}\sim 5.6$ at $t-t_{\mathrm{mrg}}\sim 4.2\,[\mathrm{ms}]$. In contrast, with HLLE at this resolution early amplification increases to $\mathcal{A}\sim 7.5$ at $t-t_{\mathrm{mrg}}\sim 5.8\,[\mathrm{ms}]$. After decay from the transient peak and stabilisation subsequent to early post-merger dynamics, we find a slow growth in $\mathcal{A}$ for $t-t_{\mathrm{mrg}} \gtrsim 10\,[\mathrm{ms}]$. We observe, at late time, a qualitatively common profile in $\mathcal{A}$ emerges and leads to $\mathcal{A}_{\mathrm{HLLE}}(65)\simeq 33 > \mathcal{A}_{\mathrm{LLF}}(65) \simeq 30$. Overall this suggests that HLLE confers an improvement over LLF for the resolution of early time amplification. Note that at merger (and in its immediate aftermath) $\mathcal{A}$ depends sensitively on $\delta x$. This behaviour is expected \cite{Kiuchi:2015sga}, as higher resolution more accurately captures the turbulent amplification that develops at the shear interface when the two stars come into contact.

To inspect directly the effect of resolution, we perform an HLLE run with finest level resolution of $\delta x_m$. The behaviour qualitatively changes and is characterised by an increased maximum ampltitude (relative to $\delta x_c$), together with an early post-merger plateau of $\mathcal{A}\in[11.5,\,17.5]$ over a window of $t-t_{\mathrm{mrg}}\in[1.5,\,7.7]\,[\mathrm{ms}]$.

In the case of runs featuring a ten-fold increase of initial magnetic field strength we again observe a compatible resolution-dependent amplification. In contrast the early post-merger plateau is suppressed for $\delta x_m$. At later times $t-t_{\mathrm{mrg}} \gtrsim 40\,[\mathrm{ms}]$, we also observe that for finest resolution runs involving $\delta x_c$, a plateau in $\mathcal{A}(t)$ appears to emerge where $\mathcal{A}\sim13.2$. 
We emphasise that these runs do not assume any underlying grid symmetry, which we have demonstrated can potentially skew (artificially enhance) the amplification factor in GRMHD only runs \cite{Gutierrez:2025gkx, Cook:2025frw}.

The divergence-free constraint $\partial{}_i[B{}^i]=0$ is well-preserved to near truncation-error level throughout the evolution, at both (initial) field strengths tested, and for both resolutions. This confirms that the extensive code-infrastructure changes involved during incorporation of neutrino transport, together with the treatment of matter field coupling, has left our underlying constrained-transport scheme intact.

\begin{figure}[t]
  \centering
    \includegraphics[width=0.49\textwidth]{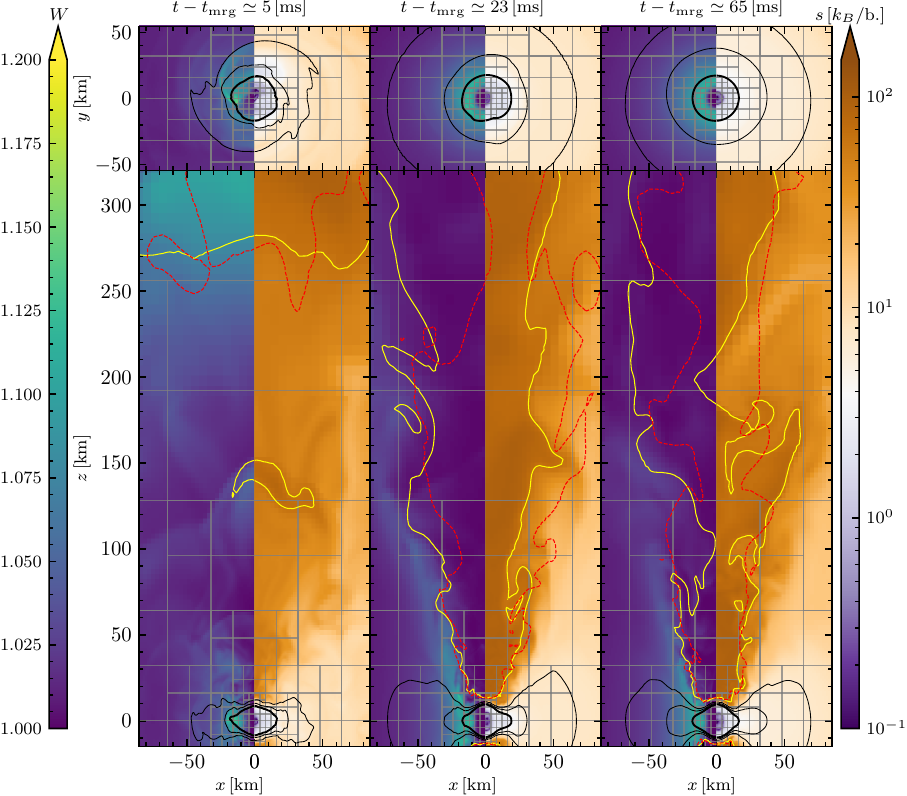}
    \caption{Snapshots during the evolution of the Lorentz factor $W$ (left half of each sub-panel) and entropy per baryon $s\,[k_B/\mathrm{b}.]$ (right side of each sub-panel) for MHD run with HLLE at finest resolution $\delta x_c$, and $\widetilde{\mathcal{B}}_{\mathrm{max}}\big|_{t=0}=5\times 10^{15}\,[\mathrm{G}]$. Isodensity contours in solid black depict $\rho\in\{10^{11},\,10^{12},\,10^{13}\}\,[\mathrm{g}\,\mathrm{cm}^{-3}]$.
        We indicate $s=65\,[k_B/\mathrm{b}.]$ with contours in yellow, and for comparison a comparable run in the absence of magnetic fields is depicted with red dashed contours. The difference between MHD and HD entropy contours is minor at this resolution, consistent with the magnetic field remaining dynamically subdominant at the resolutions accessible on the reduced grid.     }
 \label{fig:slice_xyxz_W_spb}
\end{figure}

In Fig.\ref{fig:slice_xyxz_W_spb}, we show snapshots of the Lorentz factor $W$ and entropy per baryon $s$ during the post-merger evolution of the HLLE MHD run at $\delta x_c$ with standard initial magnetic field strength. A high-entropy funnel emerges which we identify with a $s=65\,[k_B/\mathrm{b}.]$ contour. Comparison with a corresponding HD run indicates that the magnetic field remains dynamically subdominant at the resolution employed.

\begin{figure}[t]
  \centering
    \includegraphics[width=0.49\textwidth]{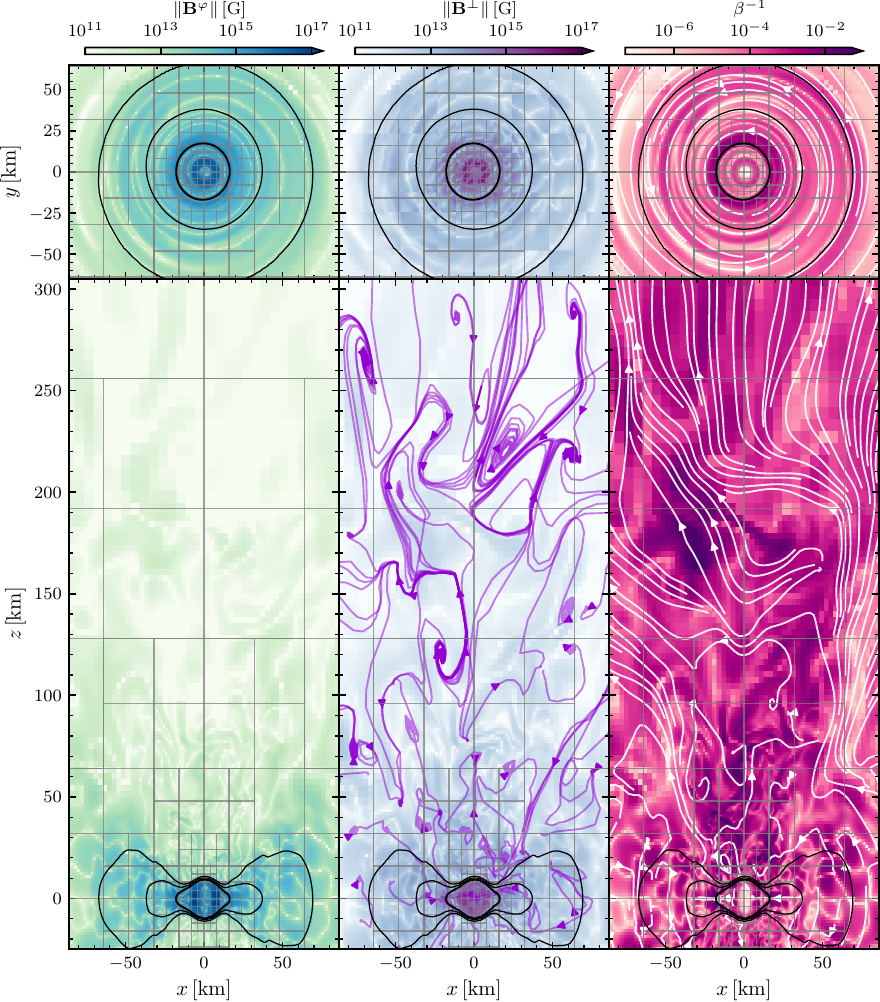}
    \caption{Azimuthal (toroidal) component of magnetic field $\Vert \mathbf{B}^\varphi\Vert $ (left-most), perpendicular (poloidal) component of magnetic field $\Vert \mathbf{B}^\perp\Vert $ with violet stream-lines showing flow of $\mathbf{B}^\perp$ super-posed (middle), and inverse plasma beta $\beta^{-1}$ with white stream-lines showing flow of $v{}^i$ super-posed (right-most), are shown at $t-t_{\mathrm{mrg}}\simeq 65\,[\mathrm{ms}]$ for HLLE run, with finest resolution of $\delta x_c$, and $\widetilde{\mathcal{B}}_{\mathrm{max}}\big|_{t=0}=5\times 10^{15}\,[\mathrm{G}]$. Isodensity contours in solid black depict $\rho\in\{10^{11},\,10^{12},\,10^{13}\}\,[\mathrm{g}\,\mathrm{cm}^{-3}]$.     }
 \label{fig:slice_xyxz_B_invbeta}
\end{figure}

Figure~\ref{fig:slice_xyxz_B_invbeta} depicts the decomposition of the magnetic field into toroidal ($\Vert\mathbf{B}^{\varphi}\Vert$) and poloidal ($\Vert\mathbf{B}^{\perp}\Vert:=\Vert\mathbf{B}-\mathbf{B}^{\varphi}\Vert$) components, together with the inverse plasma beta $\beta^{-1}$, at $t-t_{\mathrm{mrg}}\simeq 65\,[\mathrm{ms}]$.
The snapshot displays a central funnel along the polar axis has begun to form, however, due to our coarse resolution, and early time, we do not yet reach a steady (collimated) outflow regime~\cite{Musolino:2024sju}.
Achieving a self-consistent magnetically driven jet requires both higher resolution to adequately capture MRI-driven amplification and longer evolution times to allow the development of a sustained magnetically dominated outflow~\cite{Kiuchi:2023obe,Combi:2023yav}.

\subsection{Binary neutron stars: production runs}
\label{ssec:bns_prod}

For our production grids we fix the (magneto)-hydrodynamical solver to be LLF. The overall spatial extent is now selected as $x_D=2268.1\,[\mathrm{km}]$. Initial data is prepared targetting $45\,[\mathrm{km}]$ separation. We again utilize the DD2 EOS \cite{Typel:2009sy,Hempel:2009mc}. This leads to ID with NS component masses of $M_1=M_2=1.302\,[\Msun]$, and baryonic masses $M_{1b}=M_{2b}=1.350\,[\Msun]$. The initial ADM angular momentum of the BNS system is $J{}_{\mathrm{ADM}}\simeq 7.274\,[\Msun^2]$. A second physical scenario we investigate, to demonstrate stability under excision (\S\ref{sssec:sfho_excision}), is a collapsing model. For this we use the SFHo EOS \cite{Steiner:2012rk}. This leads to ID with NS component masses of $M_1=M_2=1.299\,[\Msun]$, and baryonic masses $M_{1b}=M_{2b}=1.350\,[\Msun]$. In this case, the initial BNS configuration satisfies $J{}_{\mathrm{ADM}}\simeq 7.268\,[\Msun^2]$. For both choices of EOS the respective BNS systems initially have an ADM mass of $M{}_{\mathrm{ADM}}\simeq 2.674\,[\Msun]$. For runs involving magnetic fields we introduce fields as described in \S\ref{ssec:bns_calib}.

Based on these choices, we find that the binary undergoes ${\sim}4.5$ orbits, prior to merger. In the case of DD2 EOS, based on the $\delta x_f$ run, we find that time of merger occurs at $t_{\mathrm{mrg}}\simeq 14\,[\mathrm{ms}]$. In the case of SFHo EOS, we find $t_{\mathrm{mrg}}\simeq 16\,[\mathrm{ms}]$.

\subsubsection{Scalar monitors and ejecta properties}

We begin the analysis of the production runs by examining scalar monitors and ejecta properties for the DD2 model across all three resolutions of $\mathcal{T}_\rho$, in both the MHD and HD settings.

\begin{figure}[t]
  \centering
    \includegraphics[width=0.49\textwidth]{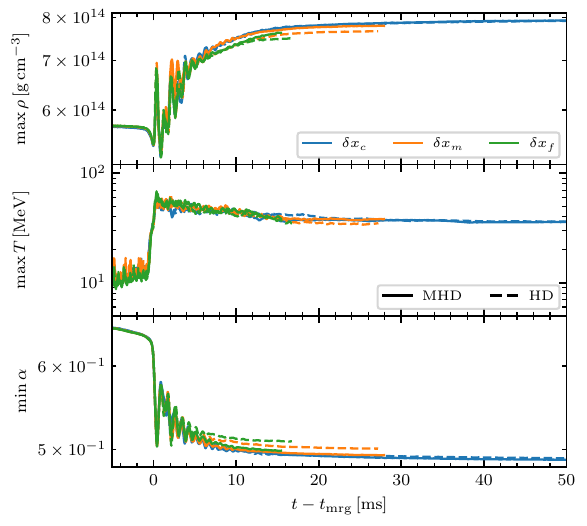}
	\caption{Scalar monitors for runs involving DD2 EOS. We show: maximum density (upper panel); maximum temperature (middle panel); minimum of the lapse (lower panel). Multiple simulations are depicted at distinct finest level resolutions, together with in the presence and absence of magnetic fields (see top and middle panels for legends).
    }
 \label{fig:DD2_full_sca_mons_A}
\end{figure}

In Fig.\ref{fig:DD2_full_sca_mons_A} we show the the evolution of the maximum density, maximum temperature, and minimum lapse for the DD2 runs, shifted to the respective time of merger for each binary. In the immediate post-merger phase, we find oscillations in the maximum density, that damp away, towards a gradual secular increase as the remnant relaxes and contracts. At $t-t_{\mathrm{mrg}}\sim 15\,[\mathrm{ms}]$ difference at the percent level emerge for $\delta x_m$, and few percent for $\delta x_f$.

The maximum temperature rises sharply towards merger and shows a peak where (resolution-dependent) values occur between $\sim 55\,[\mathrm{MeV}]$ and $68\,[\mathrm{MeV}]$ with decay thereafter as the remnant cools. While present, the differences between MHD and HD are minimal at all three resolutions for the durations of simulation time presented here.

\begin{figure}[t]
  \centering
    \includegraphics[width=0.49\textwidth]{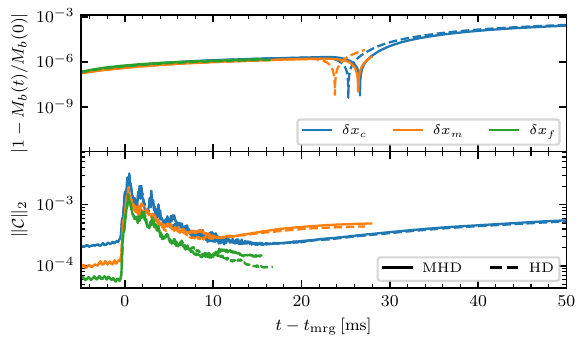}
    \caption{Scalar monitors for runs involving DD2 EOS. We show: conservation of baryonic mass (upper panel); collective constraint monitor (lower panel). Multiple simulations are depicted at distinct finest level resolutions, together with in the presence and absence of magnetic fields.}
 \label{fig:DD2_full_sca_mons_B}
\end{figure}

In Fig.\ref{fig:DD2_full_sca_mons_B} we show the relative error in baryonic mass conservation where we observe that irrespective of resolution or (M)HD it is satisfied to $\mathcal{O}(10^{-6})$ during the early post-merger phase. We attribute the zero-crossing at $\sim25\,[\mathrm{ms}]$ to mass outflow through the boundary of the computational domain. The collective constraint monitor \cite{Weyhausen:2011cg}, which serves as a collective proxy for Hamiltonian, momentum, and auxiliary Z4c field errors, is also depicted and confirms that numerical violations remains controlled throughout the evolution. Indeed a local maximum is attained immediately at merger, at the $\mathcal{O}(10^{-3})$ level, decaying away (on account of constraint damping).

The magnetic field diagnostics broadly follow the behaviour already investigated in \S\ref{sssec:mag_reduced}, albeit here we have a short duration of evolution of a run involving finest level resolution of $\delta x_f$. We find that for this latter run, a maximum amplification factor of $\mathcal{A}=33$ in the immediate aftermath of merger. We again observe that the divergence-free constraint is well-preserved to near truncation-error level.

\begin{figure}[t]
  \centering
    \includegraphics[width=0.49\textwidth]{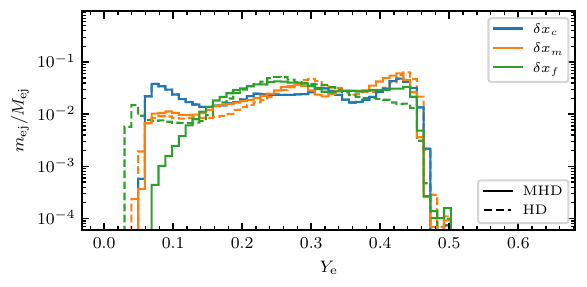}
    \includegraphics[width=0.49\textwidth]{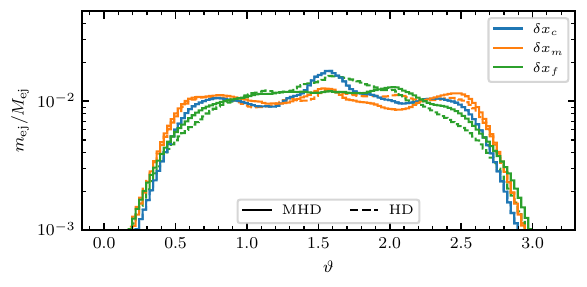}
    \caption{Histograms of ejecta involving electron fraction (upper panel), and angular distribution (lower panel).
    Resolutions $\delta x_c$, $\delta x_m$, and $\delta x_f$ are shown in blue, yellow, and green respectively. We also compare MHD (solid lines), and HD (dashed lines).
    For runs with finest resolution of $\delta x_c$ and $\delta x_m$, the cut-off time for the histogram is to $20\,[\mathrm{ms}]$ post-merger, whereas for runs with finest-resolution of $\delta x_f$ a final time of $15\,[\mathrm{ms}]$ post-merger is taken. See also Fig.\ref{fig:DD2_ej_full_grid_B}.}
 \label{fig:DD2_ej_full_grid_A}
\end{figure}

\begin{figure}[t]
  \centering
    \includegraphics[width=0.49\textwidth]{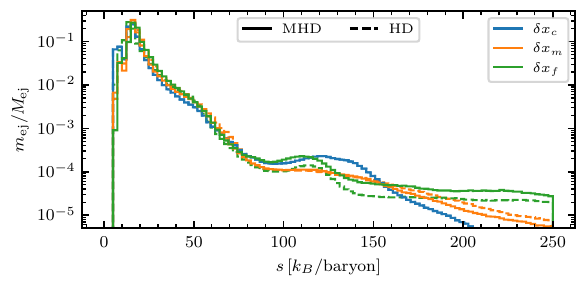}
    \includegraphics[width=0.49\textwidth]{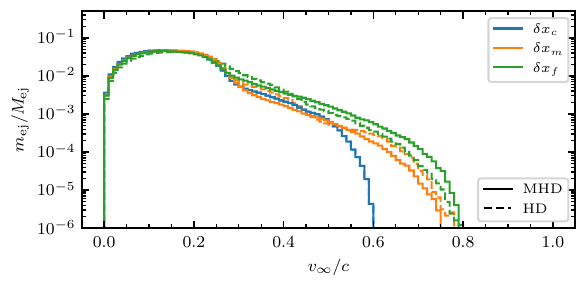}
    \caption{Histograms of ejecta involving entropy per baryon (upper panel), and asymptotic velocity (lower panel).
    Resolutions $\delta x_c$, $\delta x_m$, and $\delta x_f$ are shown in blue, yellow, and green respectively. We also compare MHD (solid lines), and HD (dashed lines).
    For runs with finest resolution of $\delta x_c$ and $\delta x_m$, the cut-off time for the histogram is to $20\,[\mathrm{ms}]$ post-merger, whereas for runs with finest-resolution of $\delta x_f$ a final time of $15\,[\mathrm{ms}]$ post-merger is taken. See also Fig.\ref{fig:DD2_ej_full_grid_A}.}
 \label{fig:DD2_ej_full_grid_B}
\end{figure}

Ejecta histograms are shown in Fig.\ref{fig:DD2_ej_full_grid_A} and Fig.\ref{fig:DD2_ej_full_grid_B}, depicting the distributions in electron fraction $Y_{\mathrm{e}}$, polar angle $\theta$, entropy per baryon $s$, and asymptotic velocity $v_\infty$ at fixed post-merger time. At $\delta x_c$ and $\delta x_m$ finest level simulations, there are only minor differences in the $Y_{\mathrm{e}}$ distribution. For $\delta x_c$ we find a broad profile with slight peaks at $Y_{\mathrm{e}}\simeq 0.08$ and $Y_{\mathrm{e}}\simeq 0.42$. In the case of $\delta x_m$ the lower end of the distribution is marginally suppressed. On the other hand when considering $\delta x_f$ we see a substantial difference between MHD and HD, wherein the presence of magnetic fields entirely suppressed the low end of the distribution function when contrasted against its hydro-only counterpart. This latter resolution-sensitive feature, has also been seen to emerge for runs (imposing bitant symmetry) in the absence of magnetic fields, at higher resolution than those employed in this work~\cite{Zappa:2022rpd}. The angular distribution is concentrated near the equatorial plane, typical of the dynamical ejecta channel~\cite{Radice:2016dwd,Combi:2022nhg}. 
In the case of the entropy distribution we find a peak at $s\sim 10$--$20\,[k_B/\mathrm{b}.]$, with a tail towards higher values corresponding to shock-heated material~\cite{Kiuchi:2022nin,Schianchi:2023uky}. For (M)HD simulations with $\delta x_f$ we observe a secondary peak at $s \sim 100\,[k_B/\mathrm{b}.]$ where we find that that the presence of magnetic fields enhances this component. Finally asymptotic velocity shows a broad peak at $v_\infty\sim 0.1$--$0.2\,[\mathrm{c}]$, followed by gradually decaying tail, with the detailed shape sensitive to resolution. In the case of $\delta x_f$ we find that the presence of magnetic fields enhances the high velocity component tail, over that of the comparable resolution HD run.

\begin{figure}[t]
  \centering
    \includegraphics[width=0.49\textwidth]{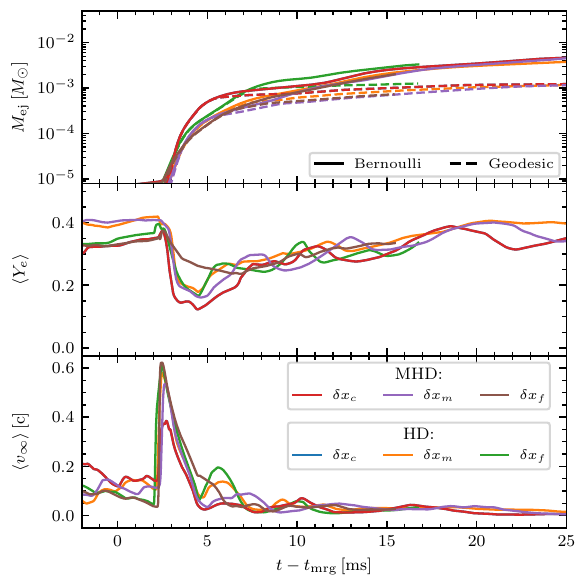}
    \caption{Ejecta for runs involving DD2 EOS. (Upper panel) Cumulative ejected mass $M_{\mathrm{ej}}$ according to Bernoulli and geodesic criteria in solid and dashed lines respectively.
    (Middle panel) Average electron fraction $\langle Y_{\mathrm{e}}\rangle$ passing through extraction sphere.
    (Lower panel) Average asymptotic velocity $\langle v_\infty \rangle$ passing through extraction sphere.
    (Common) Multiple resolutions, in the presence and absence of magnetic fields, as indicated by the lower panel insets are depicted. Extraction sphere of radius $R\simeq 295\,[\mathrm{km}]$ is utilized. See text for discussion.
        }
 \label{fig:DD2_ej_full}
\end{figure}

Cumulative ejecta mass $M_{\mathrm{ej}}$, average electron fraction $\langle Y_{\mathrm{e}}\rangle$, and average asymptotic velocity $\langle v_\infty\rangle$ are compared in Fig.\ref{fig:DD2_ej_full}.  Both the Bernoulli and geodesic unbound criteria are shown. 

In the early post-merger, we find that for simulations based on $\delta x_f$, $M_{\mathrm{ej}}$ is boosted in the absence of magnetic fields. Magnetic pressure may provide marginal additional support to the remnant, delaying mass ejection at the highest resolution where the field is best resolved.

Finally, focusing on $\langle v_\infty\rangle$ we see a resolution hierarchy emerge, whereby shortly post-merger ($t\simeq2.5\,[\mathrm{ms}]$) the simulations from highest to lowest resolution show the highest to lowest peak $\langle v_\infty \rangle$ respectively. A potential explanation for this may be related to higher resolution better capturing the sharp shock and tidal-tail interfaces responsible for accelerating dynamical ejecta, thereby producing faster outflows.

\begin{figure}[t]
  \centering
    \includegraphics[width=0.49\textwidth]{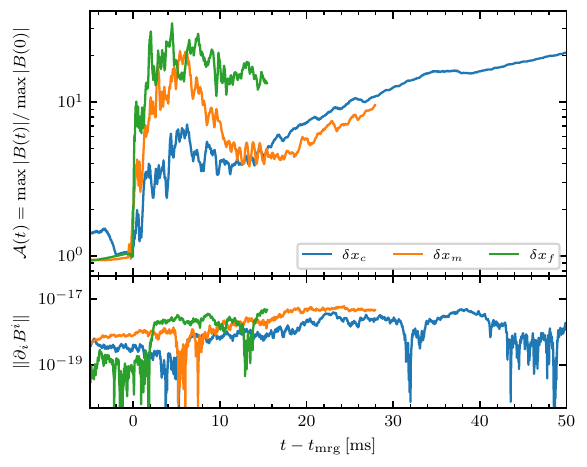}
    \caption{(Upper panel): amplification of maximum magnetic field strength relative to the maximum initial value for DD2 EOS runs at distinct resolutions, indicated by the legend. (Lower panel): Preservation of divergence-free constraint.}
 \label{fig:BfieldAmpDiv_full}
\end{figure}

\subsubsection{Neutrino luminosities and average energies}
\label{sbsbsec:nu_lum}

\begin{figure}[t]
  \centering
    \includegraphics[width=0.49\textwidth]{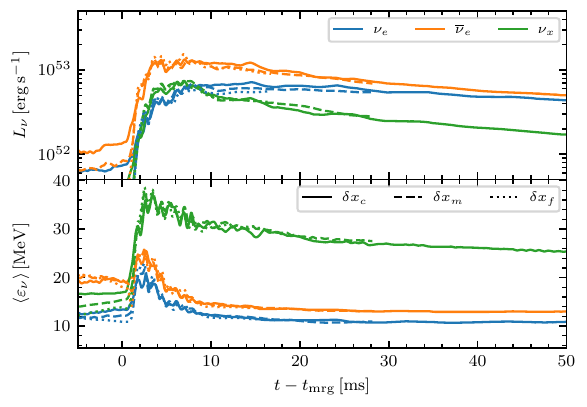}
    \caption{Neutrino luminosity $L_\nu$ (upper) and average energy $\langle \varepsilon_\nu \rangle$ (lower) at distinct resolutions, for long-lived remnant run in the absence of magnetic fields. The species $(\nu_e,\,\overline{\nu}_e,\,\nu_x)$ are coloured in blue, orange, and green respectively. In solid, dashed, and dotted lines, data corresponding to simulations with finest resolutions $\mathcal{T}_\rho$ respectively is displayed. Observe that the values of the computed luminosity remain robust over the distinct resolutions.
        Data has been smoothed using a rolling average with width of $0.5\,[\mathrm{ms}]$.}
 \label{fig:DD2_hd_lum_nrg}
\end{figure}

During BNS inspiral, our simulations point to negligible neutrino emission, as expected for matter close to beta-equilibrium. This changes during the merger-remnant phase. At merger, shock heating and remnant bounce drive a rapid temperature increase, activating intense neutrino emission. In Fig.\ref{fig:DD2_hd_lum_nrg} we display neutrino luminosities, together with average energies, for evolution of ID leading to a long-lived remnant, for each resolution of the triplet $\mathcal{T}_\rho$, in the absence of magnetic fields. Within a few milliseconds post-merger, we find that the luminosity hierarchy $L_{\nu_e}<L_{\nu_x}<L_{\overline{\nu}_e}$ develops (compatible with ~\cite{Radice:2021jtw,Zappa:2022rpd,Schianchi:2023uky}). This is followed by $~{\sim} 8\,[\mathrm{ms}]$ of small fluctuations in brightness, generated by oscillations in the remnant, followed by decay. As can be seen in the figure $\max_t L_{\overline{\nu}_e}\sim10^{53}\,[\mathrm{erg}\,\mathrm{s}^{-1}]$ which is compatible with (e.g.)~\cite{Cusinato:2021zin,Espino:2023mda}.

In a similar vein, the average neutrino energies are seen to obey the expected hierarchy $\langle \varepsilon_{\nu_e}\rangle < \langle \varepsilon_{\overline{\nu}_e}\rangle < \langle \varepsilon_{\nu_x}\rangle$ \cite{Ruffert:1998vp,Foucart:2016rxm,Cusinato:2021zin}. This is as a result of species $\nu_e$ decoupling from matter at lower densities and temperatures compared to the two other species \cite{Endrizzi:2019trv}. 
The complementary simulations of this ID, involving magnetic fields, do not lead to a significant change in $L_\nu(t)$, and $\langle \varepsilon_\nu(t)\rangle$, at the resolutions investigated.

\begin{figure}[t]
  \centering
    \includegraphics[width=0.49\textwidth]{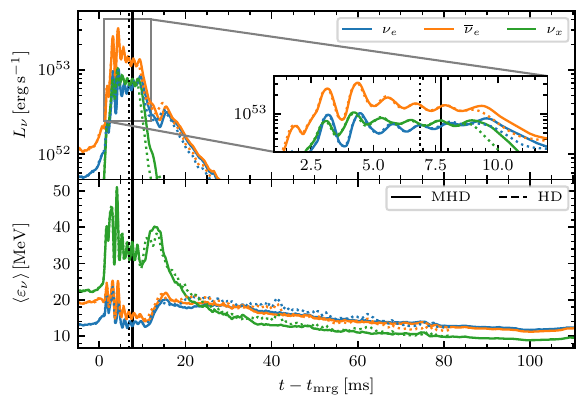}
    \caption{Neutrino luminosity $L_\nu$ (upper) and average energy $\langle \varepsilon_\nu \rangle$ (lower), for ID that results in collapse, comparing (M)HD in (solid) dotted lines. Both runs are performed with $\delta x_c$. Formation time of apparent horizon is indicated with vertical black lines. Observe that the run involving magnetic field collapses $\sim0.8\,[\mathrm{ms}]$ earlier. Decay in the magnitude of the luminosities is correspondingly shifted. Data has been smoothed using a rolling average with width of $0.5\,[\mathrm{ms}]$.}
 \label{fig:SFHo_mhd_lum_nrg}
\end{figure}

In Fig.\ref{fig:SFHo_mhd_lum_nrg} we show $L_\nu(t)$, and $\langle \varepsilon_\nu(t)\rangle$, for evolution of ID that leads to collapse. Comparison is made (at fixed resolution $\delta x_c$) with and without magnetic fields present. In the absence of magnetic field, we find $\max_t L{}_{\overline{\nu}_e}\sim 3\times 10^{53}\,[\mathrm{erg}\,\mathrm{s}^{-1}]$, in broad agreement with \cite{Espino:2023mda} (though note resolutions differ). As \cite{Espino:2023mda,Schianchi:2023uky} we confirm a higher peak luminosity of SFHo relative to the DD2. For the run with magnetic field, collapse time is advanced by $\delta t\sim 0.8\,[\mathrm{ms}]$. This impacts neutrino luminosity and average energy, where a similar time shift is present in the onset of their rapid decay. As already observed in \cite{Radice:2021jtw} (see also \cite{Zappa:2022rpd}), the behaviour of $L_\nu$ and $\langle \varepsilon_\nu\rangle$ can be quite sensitive to resolution\footnote{Direct convergence testing in neutrino luminosity for a single, non-rotating, radiating NS is already involved \cite{musolino2024practicalguidemoment}.} during the collapse phase. 

\subsubsection{Gravitational waveforms}

Consider the gravitational wave strain $h$ for the dominant $(2,2)$-mode. According to the study of~\cite{Zappa:2022rpd}, a finest level resolution of at least $\delta x \lesssim 185\,[\mathrm{m}]$ (and preferably finer) during the course of the simulation is required in order to discern microphysical effects (neutrino transport specifically) on the post-merger gravitational wave signal. At the resolutions employed in this work any transport-induced variation is likely to be within the numerical error budget (see also~\cite{Radice:2021jtw,Schianchi:2023uky}). Nonetheless, as a consistency check, we have verified that a convergence order of $\simeq 2$ in the $(2,2)$-mode phasing and amplitude to merger based on simulations involving the resolution triplet $\mathcal{T}_\rho$. Post-merger, the complex flow structure of the remnant is sensitive to resolution.  There we found approximately first-order convergence in the postmerger regime, consistent with the findings of~\cite{Zappa:2022rpd}. These verifications were performed for both HD and MHD runs.

\begin{figure}[t]
  \centering
  \includegraphics[width=0.49\textwidth]{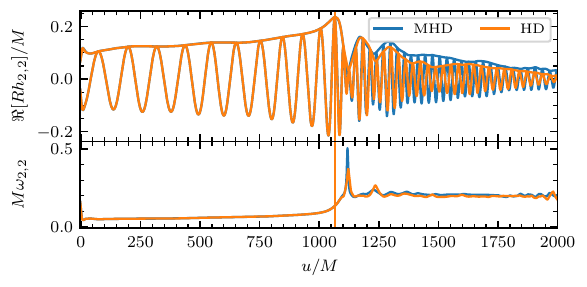}
  \caption{Gravitational wave strain $h$ for the dominant $(2,2)$-mode (upper) and instantaneous frequency (lower), extracted at $R\simeq 295\,[\mathrm{km}]$, comparing MHD (solid) and HD (dashed) for the DD2 configuration at resolution $\delta x_m$.
  }
  \label{fig:DD2_full_wvf}
\end{figure}

In Fig.\ref{fig:DD2_full_wvf} we compare MHD and HD waveforms for the DD2 configuration at $\delta x_m$.  The inspiral phase shows good agreement between the two, as expected given the negligible dynamical role of the magnetic field prior to merger. In the post-merger phase, both waveforms exhibit comparable dominant postmerger frequency and overall amplitude envelope, though the MHD run displays marginally higher amplitude in the post-merger phase, supported by additional contributions from magnetic pressure in the remnant.

\subsubsection{Robustness of excision}
\label{sssec:sfho_excision}

We now turn to the SFHo runs in order to asses the robustness of our excision strategy (see \S\ref{sbsec:ahf_excision_tap}) for handling collapse in the BNS merger context. The SFHo model, for the parameters selected within this work, leads to rapid collapse.
If we focus specifically on the case of runs that (initially) involve finest level resolutions of $\delta x_c$, then for the run involving magnetic fields the time elapsed between merger and apparent horizon (AH) formation is $7.9\,[\mathrm{ms}]$, whereas in the absence\footnote{In the HD case this duration compares favourably to a remnant life-time of $5.7\,[\mathrm{ms}]$ found in the work of \cite{Jacobi:2025eak}.} of magnetic fields we find $8.7\,[\mathrm{ms}]$.

As observed during the rotating collapse simulations described in \S\ref{sbsec:rot_collapse}, utilizing $\delta x_c$ may be insufficient to adequately resolve a developing puncture. Indeed there increased resolution plays an important role in sharpening the consistency of properties associated with the resultant AH. Motivated by this, together with a requirement for long-term stability, we adopt a simple strategy of doubling the resolution at first time of AH detection (denoted $t_{\mathrm{AH}}$). In particular, over $\mathcal{B}_r(t;\,\mathbf{c})$, for $t\geq t_{\mathrm{AH}}$,  we add a level of refinement such that $\delta x_c \leftarrow \delta x_c/2$. This increases the finest-level resolution to $184.6\,[\mathrm{m}]$ effectively setting it to $\delta x_f$. Simultaneously, we also double the CFL to $0.5$ so as to not incur a halving in the evolution speed.

\begin{figure*}[t]
  \centering
  \includegraphics[width=\textwidth]{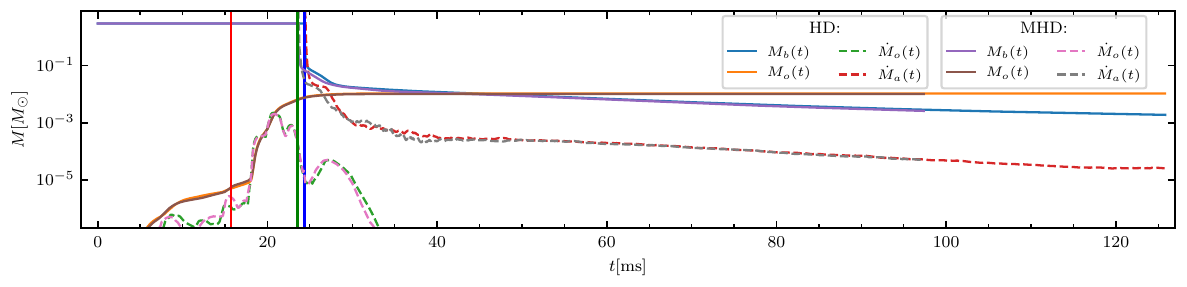}
  \caption{Mass as a function of time for simulations involving SFHo EOS and horizon-based excision. Here $\delta x_c$ is the finest resolution prior to AH formation whereupon one level of refinement is added. The baryonic mass $M_b$, outflow mass $M_o$, together the rate of change $\dot{M}_o$, and rate of accretion $\dot{M}_a$ are depicted in the case of (M)HD (see legend). Vertical red line depicts merger time $t\simeq15.7\,[\mathrm{ms}]$. Time of AH formation for MHD is $t_{\mathrm{AH}}\simeq23.6\,[\mathrm{ms}]$ and indicated by vertical green line. Similarly for HD $t_{\mathrm{AH}}\simeq24.4\,[\mathrm{ms}]$ and indicated by the vertical blue line.  We observe long-term stability beyond AH formation in both cases.}
  \label{fig:SFHo_outflow}
\end{figure*}

The result of employing the aforementioned strategy is shown in Fig.\ref{fig:SFHo_outflow}. We observe that upon formation of the AH, addition of the refinement level, and activation of the tapered excision, the baryonic mass ($M_b$) decreases monotonically as the remnant disc accretes.
The outflow mass $M_o$ saturates and the corresponding mass outflow rate $\dot{M}_o$ drops to negligible values within ${\sim} 10\,[\mathrm{ms}]$ of horizon formation. This clearly demonstrates that the excision procedure does not introduce spurious mass injection into the computational domain. Both MHD and HD runs remain stable for the full duration of the simulation ${\sim} 100 \,[\mathrm{ms}]$ and ${\sim} 125 \,[\mathrm{ms}]$ respectively (${\sim} 75 \,[\mathrm{ms}]$ and ${\sim} 100 \,[\mathrm{ms}]$ post-AH formation), with the accretion rate $\dot{M}_a$ following a gradual, declining trend. At late times, the magnetic field studied here does not appear to have a significant impact on $\dot{M}_a$. The neutrino luminosities for these runs, discussed in \S\ref{sbsbsec:nu_lum} and shown in Fig.\ref{fig:SFHo_mhd_lum_nrg}, decay rapidly after collapse, as expected once the hot remnant is swallowed by the horizon. The $0.8\,[\mathrm{ms}]$ shift in onset of decay of the luminosities corresponds well with the earlier collapse of the run involving magnetic fields.

We remark that the number of active points over the diameter of tapering region is $\sim17$ where the strength of equation source-term damping is modulated point-wise by the magnitude of the taper function, as controlled by how its parameters are chosen (\S\ref{sbsec:ahf_excision_tap}).

A well-known potential difficulty associated with methods that rely on moving-punctures is the coordinate drift of the puncture location~\cite{Brugmann:2008zz}. This can cause an AH (and corresponding excised region) to migrate across the grid. In part, this effect can be damped through judicious modification of the gauge (shift-condition)~\cite{Brugmann:2008zz,Most:2021ytn}, and alternative gauge choices~\cite{Shibata:2021xmo} (see also the discussion in~\cite{Radice:2023zlw}).
In our simulations we observe a modest puncture drift wherein the remnant, at the final times we evolve to has $\mathrm{argmin}_{\mathbf{x}^*\in\Omega}\alpha(\mathbf{x}^*)$ such that $\Vert \mathbf{x}^*\Vert\simeq3\,[\mathrm{km}]$ in both MHD and HD runs. This does not appear to pose issues for stability, which may potentially be explained by our dynamical trackers allowing for the position of the AH to adapt in time. On the other hand, this dynamical drift may be undesirable in light of physical modelling considerations.
We leave investigation of the effect of drift-correction strategies to future work.

The question of how to treat matter and radiation fields inside the apparent horizon in the presence of neutrino transport has also recently been investigated in the context of stellar core collapse \cite{kuroda2023failedsupernovasimulations,kuroda2024numericalrelativitysimulations}. While a different physical scenario, we may nonetheless briefly compare the excision strategy. There simulations are performed in axiymmetry, using full numerical relativity (BSSN). 
In their approach, metric variables are evolved everywhere with moving-puncture gauge, while matter and radiation fields are (strictly) excised inside $0.5\,r_\mathrm{AH}$. To avoid the introduction of spurious oscillations, that could propagate outward, a buffer zone between excision surface and AH is essential for stability.  Our approach differs in that the tapering function described in \S\ref{sbsec:ahf_excision_tap} modifies state vectors only through the evolution, on the interior of the AH, driving them towards atmosphere values without requiring a buffer zone. Despite different design choices, both approaches lead to stable evolution, post-collapse. We emphasise that the simulations of~\cite{kuroda2023failedsupernovasimulations,kuroda2024numericalrelativitysimulations} are restricted to axisymmetry, whereas the present work demonstrates excision robustness in full 3D binary neutron star mergers with M1 transport. 

\section{Conclusion}
\label{sec:conclusion}

We have presented the implementation and a robustness validation of an energy-integrated, moment-based (M1+N0) neutrino-transport scheme within the general-relativistic radiation magnetohydrodynamics code \GRAthena{}.
The transport formulation is directly inspired by the approach of~\cite{Radice:2021jtw}, adapted to the
cell-centered adaptive mesh refinement (AMR) infrastructure we have previously developed in \cite{Daszuta:2021ecf,%
Cook:2023bag,%
Daszuta:2024ucu}.
The coupled GR(M)HD-M1+N0 system is evolved via a split-step method: the choice of integrator for the GR(M)HD sector is flexible,
whereas the radiation-matter coupling is treated with a semi-implicit scheme.
Motivated by parallel developments in \texttt{AthenaK} \cite{Fields:2024pob}, a low-order flux-correction method, is now also implemented in \GRAthena{} during treatment of the hydrodynamical evolution. Conservative AMR with level-to-level flux corrections improves global conservation of salient quantities across refinement boundaries. For the M1+N0 subsystem a high-order, asymptotic-preserving scheme is hybridized with the
Local-Lax-Friedrichs (LLF) prescription for the numerical fluxes. We also introduced a simple flux correction strategy for this subsystem to further improve robustness. Closure of the moment scheme is handled via a standard, parametrized interpolatory approach. In order to handle gravitational collapse we introduced a novel excision method (\S\ref{sbsec:ahf_excision_tap}) based on (smooth) tapering, confined to apparent horizon (AH) interiors, which leverages our online AH finder.

We demonstrated robustness in an extensive suite of standard test problems (\S\ref{sec:num_test}) that probe complementary physical regimes typically encountered during (e.g.)~BNS merger simulations. In particular: optically thin advection, diffusion of a top-hat profile, diffusion in a moving medium, shadow casting, a radiating homogeneous sphere, and photon propagation in curved spacetime (gravitational light bending). A cross-code opacity comparison with the \THC{} code~\cite{Radice:2021jtw} confirms that the ported \texttt{weakrates} opacity library, together with the equilibrium handling and fluid-sector coupling, reproduces expected results.

As a first application, we studied the gravitational collapse of a cold, uniformly rotating, magnetized neutron star(\S\ref{sbsec:rot_collapse}).  This scenario serves as a controlled, stringent, test-bed for the novel excision procedure in the presence of radiation. The irreducible horizon mass and dimensionless spin saturate to values consistent with the initial data, and no secular drift observed at the two highest resolutions considered. In this latter resolution regime, the post AH formation neutrino luminosity decay profiles appear consistent. We also verified that the neutrino fluxes point exclusively into the horizon, in its predicted immediate vicinity. To our knowledge this constitutes a first demonstration that a
tapering-based excision strategy, yields stable and physically consistent radiation evolution during and beyond horizon formation in three spatial dimensions, without imposition of symmetry.

Subsequently, we applied our full GR(M)HD-M1+N0 framework to binary neutron star mergers (\S\ref{ssec:bns_calib},\,\S\ref{ssec:bns_prod}). This demonstrated the use of two finite-temperature equations of state (DD2 and SFHo), changes to the matter-sector Riemann solver (LLF vs HLLE), presence vs absence of magnetic fields (HD vs MHD), and variation of the extent of the computational domain, together with the finest grid resolution.
The comparison of LLF and HLLE solvers in the context of full GR(M)HD binary mergers that involve microphysics is, to our knowledge, novel: HLLE produces sharper contact features and a more compact remnant, whereas LLF is more dissipative but yields smoother profiles that can be advantageous for robustness. Magnetic fields, at the resolutions and time-scales considered here, do not appear to drastically alter the large-scale merger morphology, neutrino luminosities, or gravitational waveforms. This is consistent with the expectation that magnetically driven effects become dominant only at higher resolutions capable of capturing MRI-driven amplification~\cite{Kiuchi:2023obe}. The combination of general-relativistic MHD with M1 neutrino transport in binary neutron star simulations remains relatively unexplored: to date, only \cite{Musolino:2024sju} has performed GRMHD+M1 mergers with an M1 treatment that incorporates the (stiff) source-terms fully. The present work adds to this small but growing body of simulations and extends it by incorporating AH excision, enabling continued evolution well beyond gravitational collapse.

The robustness of the excision scheme has been demonstrated for an SFHo binary, which undergoes collapse at $\mathcal{O}(10)\,[\mathrm{ms}]$ post-merger. The system is stably evolved post-AH formation for ${\sim} 75\,[\mathrm{ms}]$ and ${\sim} 100\,[\mathrm{ms}]$ in both presence and absence of magnetic fields respectively. No spurious mass injection is observed.
Our tapering-based strategy is set to operate entirely inside the AH and has now been validated in full 3D binary mergers with M1 transport. 
A couple of caveats should be kept in mind: The grey (energy-integrated) M1 scheme employed here is known to potentially bias the neutrino mean energies and peak luminosities relative to spectral (multigroup) treatments~\cite{Cheong:2024buu}, and the single-energy approximation precludes capturing spectral features that influence nucleosynthesis-relevant quantities such as the electron-fraction distribution of the ejecta.
Ejecta properties are also sensitive to resolution~\cite{Zappa:2022rpd,Radice:2023zlw} and thus while our results are sufficient to demonstrate code robustness, quantitative analysis may require data at higher resolution. To such a future aim, our underlying infrastructure permits efficient scaling in this regard.

Two improvements are planned for the near term. First, the neutrino interaction rates will be upgraded to the \texttt{bnsnurates} library~\cite{Chiesa:2024lnu}, which provides a more complete and accurate set of weak-interaction processes.
Second, in the spirit of ~\cite{Cheong:2024buu}, the extension to a multigroup (spectral) M1 scheme is under development. This will address the systematic biases of the grey treatment and enable more reliable predictions of ejecta composition and kilonova light curves.

\begin{acknowledgments}
  BD and SB acknowledges support by the EU Horizon under ERC Consolidator Grant,
  no. InspiReM-101043372.
  EG and DR were supported by the National Science Foundation under Grant PHY-2407681.
  PH was supported by the National Science Foundation under Grant PHY-2116686.
  DR also acknowledges support from the Sloan Foundation, from the
  Department of Energy, Office of Science, Division of Nuclear Physics
  under Awards Number DOE DE-SC0021177 and DE-SC0024388, from the
  National Science Foundation under Grants No. PHY-2020275 and
  PHY-2512802.
    The authors are indebted to A.~Celentano's PRİSENCÓLİNENSİNÁİNCIÚSOL.

  Simulations were performed on the national HPE Apollo Hawk (Hunter)
  at the High Performance Computing Center Stuttgart (HLRS).
  The authors acknowledge HLRS for funding this project by providing access
  to the supercomputer HPE Apollo Hawk (Hunter) under the grant numbers INTRHYGUE/44215
  and MAGNETIST/44288.
  Simulations were also performed on SuperMUC\_NG at the
  Leibniz-Rechenzentrum (LRZ) Munich.
  The authors acknowledge the Gauss Centre for Supercomputing
  e.V. (\url{www.gauss-centre.eu}) for funding this project by providing
  computing time on the GCS Supercomputer SuperMUC-NG at LRZ
  (allocations {\tt pn67xo}, {\tt pn76li}, {\tt pn68wi} and {\tt pn36jo}).
  The numerical simulations were performed on TACC's Frontera (NSF LRAC
  allocation PHY23001) and on Perlmutter using NERSC award ERCAP0031370.
  This research used resources
  of the National Energy Research Scientific Computing Center, a DOE
  Office of Science User Facility supported by the Office of Science of
  the U.S.~Department of Energy under Contract No.~DE-AC02-05CH11231.
  Postprocessing and development runs were performed on the ARA cluster
  at Friedrich Schiller University Jena.
  The ARA cluster is funded in part by DFG grants INST
  275/334-1 FUGG and INST 275/363-1 FUGG, and ERC Starting Grant, grant
  agreement no. BinGraSp-714626.
\end{acknowledgments}

\appendix
\section{Source term derivatives}
\label{sec:appendix_jacobian}
When the radiation pressure tensor is assumed to be of interpolatory form as in \S\ref{sbsbsec:interp_lim}, (formal) derivatives of the collisional sources appearing in the Jacobian of \S\ref{sbsbsec:coll_src} may be written down as (see also \cite{Radice:2021jtw,Izquierdo:2022eaz}):
\begin{equation}
\begin{aligned}
  \frac{\delta \widetilde{\mathcal{S}}{}_1}{\delta \widetilde{E}}
  &=
  \alpha W\left(
    \kappa{}_{\mathrm{s}} \frac{\delta \widetilde{J}}{\delta \widetilde{E}}
    -\kappa{}_{\mathrm{as}}
  \right), \\
  \frac{\delta \widetilde{\mathcal{S}}{}_1}{\delta \widetilde{F}{}_j}
  &=
  \alpha W \left(
    \kappa{}_{\mathrm{s}} \frac{\delta \widetilde{J}}{\delta \widetilde{F}{}_j} +
    \kappa{}_{\mathrm{as}} v{}^j
  \right), \\
  \frac{\delta \widetilde{\mathcal{S}}{}_{1+i}}{\delta \widetilde{E}}
  &=
  -\alpha\left(
    W \kappa{}_{\mathrm{a}}
    \frac{\delta \widetilde{J}}{\delta \widetilde{E}} v{}_i +
    \kappa{}_{\mathrm{as}} \frac{\delta}{\delta \widetilde{E}} \left[
      {}^E\mathcal{P}{}^b_i \widetilde{H}{}_b
    \right]
  \right), \\
  \frac{\delta \widetilde{\mathcal{S}}{}_{1+i}}{\delta \widetilde{F}{}_j}
  &=
  -\alpha\left(
    W \kappa{}_{\mathrm{a}}
    \frac{\delta \widetilde{J}}{\delta \widetilde{F}{}_j} v{}_i +
    \kappa{}_{\mathrm{as}} \frac{\delta}{\delta \widetilde{F}{}_j} \left[
      {}^E\mathcal{P}{}^b_i \widetilde{H}{}_b
    \right]
  \right);
\end{aligned}
\end{equation}
where we used Eq.\eqref{eq:srcs_expanded} together with Eq.\eqref{eq:dec_radflux}. The neutrino current component $\widetilde{\mathcal{S}}{}_0$ is not considered here as this is solved for directly instead. Based on factorization of fiducial frame terms as in Eq.\eqref{eq:factorframe} we find:
\begin{widetext}
\begin{equation}
\begin{aligned}
  \frac{\delta \widetilde{J}}{\delta \widetilde{E}}
  &=
  W^2 +
  W^2 \frac{\left(\widetilde{F}{}_j v{}^j\right)^2}{\Vert \widetilde{F}\Vert^2} d{}_{\mathrm{t}} +
  \left(
    3-W^2-\frac{6}{2W^2+1}
  \right) d{}_{\mathrm{T}},\\
  \frac{\delta \widetilde{J}}{\delta \widetilde{F}{}_j}
  &=
  2W^2\left(
    -v{}^j
    +\frac{\widetilde{E}}{\Vert \widetilde{F} \Vert^4}\left[
      \Vert \widetilde{F} \Vert^2 v{}^j - \widetilde{F}{}_k v{}^k \widetilde{F}{}^j
    \right] \widetilde{F}{}_i v{}^i d{}_{\mathrm{t}}
    + \frac{2}{2W^2+1}\left[W^2-1\right] v{}^j d_{\mathrm{T}}
  \right)
  ,\\
  \frac{\delta}{\delta \widetilde{E}} \left[
    {}^E\mathcal{P}{}^b_i \widetilde{H}{}_b
  \right]
  & =
  W^3 \left(
    -1 - \frac{\left(\widetilde{F}{}_j v{}^j\right)^2}{\Vert \widetilde{F} \Vert^2} d_{\mathrm{t}}
    +\frac{2W^2-3}{2W^2+1} d{}_{\mathrm{T}}
  \right)v{}_i
  -\frac{W \widetilde{F}{}_k v{}^k d{}_{\mathrm{t}}}{\Vert \widetilde{F} \Vert^2}  \widetilde{F}{}_i,\\
  \frac{\delta}{\delta \widetilde{F}{}_j} \left[
    {}^E\mathcal{P}{}^b_i \widetilde{H}{}_b
  \right]
  & =
  2W^3 \left(
    v{}_i v{}^j + \left[
      \frac{\widetilde{E}\left(\widetilde{F}{}_k v{}^k\right)^2}{\Vert \widetilde{F} \Vert^4} v{}_i \widetilde{F}{}^j
      - \frac{\widetilde{E}\widetilde{F}{}_k v{}^k}{\Vert \widetilde{F} \Vert^2} v{}_i v{}^j
    \right]d{}_{\mathrm{t}}
    + v{}_i v{}^j d_{\mathrm{T}}
  \right)\\
  & \hphantom{=}
  + W\left(
    \delta{}^j_i +
    \left[
      \frac{\widetilde{E}}{\Vert \widetilde{F}\Vert^2}
      \left(
        \frac{2}{\Vert \widetilde{F} \Vert^2} \widetilde{F}{}_k v{}^k \widetilde{F}{}_i \widetilde{F}{}^j
        - \widetilde{F}{}_k v{}^k \delta{}^j_i
        - \widetilde{F}{}_i v{}^j
      \right)
    \right] d_{\mathrm{t}}
    +\delta{}^j_i d{}_{\mathrm{T}}
  \right).
\end{aligned}
\end{equation}
\end{widetext}

\end{document}